%% file: main_new.tex
\documentclass[acmsmall,noacm]{acmart}

\newif\ifTECHREPORT
\TECHREPORTtrue
\newcommand{\techref}[2]{\ifTECHREPORT{#1}\else{#2}\fi}

\acmDOI{10.1145/3729287}
\acmYear{2025}
\acmJournal{PACMPL}
\acmVolume{9}
\acmNumber{PLDI}
\acmArticle{184}
\acmMonth{6}
\received{2024-11-15}
\received[accepted]{2025-03-06}

\usepackage[utf8]{inputenc}
\usepackage[T1]{fontenc}
\usepackage{microtype}

\input{shortcut}

\begin{document}

\title{Program Skeletons for Automated Program Translation}

\author{Bo Wang}
\authornote{Both authors contributed equally to this research.}
\orcid{0000-0003-1444-0237}
\affiliation{%
  \institution{National University of Singapore}
  \city{Singapore}
  \country{Singapore}
}
\email{bo_wang@u.nus.edu}

\author{Tianyu Li}
\authornotemark[1]
\orcid{0009-0001-6180-9060}
\affiliation{%
  \institution{National University of Singapore}
  \city{Singapore}
  \country{Singapore}
}
\email{tianyuli@u.nus.edu}

\author{Ruishi Li}
\orcid{0000-0003-2513-1704}
\affiliation{%
  \institution{National University of Singapore}
  \city{Singapore}
  \country{Singapore}
}
\email{liruishi@u.nus.edu}

\author{Umang Mathur}
\orcid{0000-0002-7610-0660}
\affiliation{%
  \institution{National University of Singapore}
  \city{Singapore}
  \country{Singapore}
}
\email{umathur@comp.nus.edu.sg}

\author{Prateek Saxena}
\orcid{0000-0002-1875-8675}
\affiliation{%
  \institution{National University of Singapore}
  \city{Singapore}
  \country{Singapore}
}
\email{prateeks@comp.nus.edu.sg}

\input{sections/0abstract}



\maketitle

\input{sections/new-intro}

\input{sections/2problemoverview}

\input{sections/3decooverview}

\input{sections/4_1_syntax}

\input{sections/4_2_semantics}

\input{sections/4_3_synth}

\input{sections/6impl}

\input{sections/7eval}

\input{sections/8futurework}
\input{sections/8relatedwork}

\input{sections/Xconclusion}

\begin{acks}
\todoadd{We thank the anonymous reviewers. This research is supported by grants given by the Ministry of Education in Singapore (Tier 2 grant is MOE-T2EP20124-0007) and Cisco Research (Cisco University Research Program Fund, Silicon Valley Community Foundation).}
\end{acks}

\bibliographystyle{ACM-Reference-Format}
\bibliography{paper}

\techref{
\clearpage
\appendix

\input{sections/new-A_model}

}{}

\end{document}

%% file: shortcut.tex
\usepackage{amsmath,amsfonts}
\usepackage{array}
\usepackage{verbatimbox}
\usepackage[noend]{algpseudocode}
\usepackage{graphicx}
\usepackage{textcomp}
\usepackage{xspace}
\usepackage{xcolor}
\usepackage{listings}
\usepackage{booktabs}
\usepackage{algorithm}
\usepackage{tabularx}
\usepackage{multicol}
\usepackage{multirow}
\usepackage{soul}
\usepackage{tikz}
\usepackage{comment}
\usepackage{colortbl} %
\usepackage[breakable]{tcolorbox}

\definecolor{mygray}{HTML}{e3e6e8}
\definecolor{darkgreen}{RGB}{0, 150, 0}
\definecolor{darkblue}{RGB}{0, 0, 150}
\definecolor{cyanHighlight}{HTML}{5CFFFF}
\definecolor{yellowHighlight}{HTML}{FBFF00}
\definecolor{lightgreen}{RGB}{198,243,196}

\newcommand{\pybold}{{\sf \textbf{Python}}\xspace}
\newcommand{\py}{{\sf Python}\xspace}
\newcommand{\js}{{\sf JavaScript}\xspace}
\newcommand{\name}{\textsc{Skel}\xspace}
\newcommand{\decocr}{{\sf SkelCR}\xspace}
\newcommand{\procint}{\textcolor{darkblue}{\sf ProcEmu}\xspace}
\newcommand{\procintbold}{\textcolor{darkblue}{\sf \textbf{ProcEmu}}\xspace}

\newcommand{\jitt}{\textsc{Eot}\xspace}

\lstdefinelanguage{JavaScript}{
  keywords={import, break, case, catch, continue, debugger, default, delete, do, else, finally, for, function, if, in, instanceof, new, return, switch, throw, try, typeof, var, void, while, with, let, const},
  morecomment=[l]{//},
  morecomment=[s]{/*}{*/},
  morestring=[b]',
  morestring=[b]"
}

\lstdefinestyle{diff}{
  language=JavaScript,
  moredelim=[is][\color{red}]{--}{--},
  moredelim=[is][\color{darkgreen}]{++}{++},
}

\newcommand{\code}[1]{{\setlength\fboxsep{1pt}\colorbox{mygray}{\lstinline|#1|}}}

\newcommand{\highlightorange}[1]{{\setlength\fboxsep{1pt}\colorbox{orange}{\lstinline|#1|}}}

\newcommand{\highlightgreen}[1]{{\setlength\fboxsep{1pt}\colorbox{lime!30}{\lstinline|#1|}}}

\newcommand{\highlightlightgreen}[1]{{\setlength\fboxsep{1pt}\colorbox{lightgreen}{\lstinline|#1|}}}

\newcommand{\nobgcode}[1]{{\setlength\fboxsep{1pt}{\lstinline|#1|}}}

\newcommand{\graytext}[1]{{\setlength\fboxsep{1pt}\colorbox{mygray}{#1}}}

\definecolor{deletecolor}{rgb}{0.66, 0, 0}
\definecolor{addcolor}{rgb}{0, 0.57, 0}
\newcommand\tododel[1]{}
\newcommand\todoadd[1]{#1} 
\newcommand\ignore[1]{}

\newenvironment{myparagraph}[1]{\vspace{0.1in}\noindent{\bf #1.}}{}
\newcommand\myarrow[1]{\xrightarrow{\sf #1}}

\theoremstyle{definition}

\newcommand{\lang}{\mathcal{L}}
\newcommand{\skl}{K}
\newcommand{\sklsem}{{\hat{K}}}
\newcommand{\holes}[1]{{\sf Holes}(#1)}
\newcommand{\set}[1]{\{#1\}}

\newcommand{\cmpl}{\Gamma}

\newcommand{\pgm}{\ell}
\newcommand{\frag}{g}
\newcommand{\lsemreq}{\varphi}
\newcommand{\src}{{\sf src}}

\newcommand{\tgt}{{\sf tgt}}
\newcommand{\eq}{\sim}
\newcommand{\eqsrctgt}{{\sf Eq}_{(\src, \tgt)}}

\newcommand{\fragsrc}{\frag^\src}

\newcommand{\FLAGNEWDATA}{}

\ifdefined\FLAGNEWDATA
\input{shortdata_new}

\else

\fi

%% file: shortdata_new.tex
\newcommand{\progsizeJJJtransmapmin}{121}
\newcommand{\progsizeJJJtransmapmax}{882}

\newcommand{\percentskelauto}{\autogptfourpctJJJTOTAL}

\newcommand{\skelfourAutoProgCount}{4}
\newcommand{\skelthreeAutoProgCount}{1}

\newcommand{\gptfourbaseFailProgCount}{2}
\newcommand{\transcryptbasePassProgCount}{2}
\newcommand{\transcryptbaseNAProgCount}{5}

\newcommand{\userfixRatioSkelVSBase}{1/4}

\newcommand{\ablationGapSpecMinusChkrfn}{38}

\newcommand{\progsizeLocJJJcolorsys}{121}
\newcommand{\progsizeLocJJJheapq}{189}
\newcommand{\progsizeLocJJJhtml}{882}
\newcommand{\progsizeLocJJJmathgen}{736}
\newcommand{\progsizeLocJJJstrsim}{686}
\newcommand{\progsizeLocJJJbsdrec}{250}
\newcommand{\progsizeLocJJJrbtree}{487}
\newcommand{\progsizeLocJJJtoml}{1272}
\newcommand{\progsizeLocJJJpyevtx}{2400}

\newcommand{\progsizeSlocJJJcolorsys}{120}
\newcommand{\progsizeSlocJJJheapq}{189}
\newcommand{\progsizeSlocJJJhtml}{684}
\newcommand{\progsizeSlocJJJmathgen}{735}
\newcommand{\progsizeSlocJJJstrsim}{654}
\newcommand{\progsizeSlocJJJbsdrec}{123}
\newcommand{\progsizeSlocJJJrbtree}{366}
\newcommand{\progsizeSlocJJJtoml}{1206}
\newcommand{\progsizeSlocJJJpyevtx}{1711}

\newcommand{\hcgJJJcolorsys}{2}
\newcommand{\hcgJJJheapq}{4}
\newcommand{\hcgJJJhtml}{5}
\newcommand{\hcgJJJmathgen}{2}
\newcommand{\hcgJJJstrsim}{3}
\newcommand{\hcgJJJbsdrec}{4}
\newcommand{\hcgJJJrbtree}{5}
\newcommand{\hcgJJJtoml}{8}
\newcommand{\hcgJJJpyevtx}{26}

\newcommand{\covJJJcolorsys}{100}
\newcommand{\covJJJheapq}{100}
\newcommand{\covJJJhtml}{85}
\newcommand{\covJJJmathgen}{100}
\newcommand{\covJJJstrsim}{91}
\newcommand{\covJJJbsdrec}{100}
\newcommand{\covJJJrbtree}{89}
\newcommand{\covJJJtoml}{80}
\newcommand{\covJJJpyevtx}{72}

\newcommand{\locfJJJcolorsys}{16}
\newcommand{\locfJJJheapq}{9}
\newcommand{\locfJJJhtml}{14}
\newcommand{\locfJJJmathgen}{9}
\newcommand{\locfJJJstrsim}{10}
\newcommand{\locfJJJbsdrec}{15}
\newcommand{\locfJJJrbtree}{14}
\newcommand{\locfJJJtoml}{17}
\newcommand{\locfJJJpyevtx}{6}

\newcommand{\countfJJJcolorsys}{9}
\newcommand{\countfJJJheapq}{24}
\newcommand{\countfJJJhtml}{42}
\newcommand{\countfJJJmathgen}{82}
\newcommand{\countfJJJstrsim}{50}
\newcommand{\countfJJJbsdrec}{21}
\newcommand{\countfJJJrbtree}{27}
\newcommand{\countfJJJtoml}{47}
\newcommand{\countfJJJpyevtx}{164}
\newcommand{\countfJJJTOTAL}{466}

\newcommand{\autogptfourJJJcolorsys}{9}
\newcommand{\autogptfourJJJheapq}{21}
\newcommand{\autogptfourJJJhtml}{40}
\newcommand{\autogptfourJJJmathgen}{78}
\newcommand{\autogptfourJJJstrsim}{50}
\newcommand{\autogptfourJJJbsdrec}{21}
\newcommand{\autogptfourJJJrbtree}{27}
\newcommand{\autogptfourJJJtoml}{37}
\newcommand{\autogptfourJJJpyevtx}{160}
\newcommand{\autogptfourJJJTOTAL}{443}

\newcommand{\autogptfourpctJJJcolorsys}{100}
\newcommand{\autogptfourpctJJJheapq}{88}
\newcommand{\autogptfourpctJJJhtml}{95}
\newcommand{\autogptfourpctJJJmathgen}{95}
\newcommand{\autogptfourpctJJJstrsim}{100}
\newcommand{\autogptfourpctJJJbsdrec}{100}
\newcommand{\autogptfourpctJJJrbtree}{100}
\newcommand{\autogptfourpctJJJtoml}{79}
\newcommand{\autogptfourpctJJJpyevtx}{98}
\newcommand{\autogptfourpctJJJTOTAL}{95}

\newcommand{\fixgptfourJJJcolorsys}{0}
\newcommand{\fixgptfourJJJheapq}{3}
\newcommand{\fixgptfourJJJhtml}{2}
\newcommand{\fixgptfourJJJmathgen}{4}
\newcommand{\fixgptfourJJJstrsim}{0}
\newcommand{\fixgptfourJJJbsdrec}{0}
\newcommand{\fixgptfourJJJrbtree}{0}
\newcommand{\fixgptfourJJJtoml}{10}
\newcommand{\fixgptfourJJJpyevtx}{4}
\newcommand{\fixgptfourJJJTOTAL}{23}

\newcommand{\fixgptfourpctJJJcolorsys}{0}
\newcommand{\fixgptfourpctJJJheapq}{12}
\newcommand{\fixgptfourpctJJJhtml}{5}
\newcommand{\fixgptfourpctJJJmathgen}{5}
\newcommand{\fixgptfourpctJJJstrsim}{0}
\newcommand{\fixgptfourpctJJJbsdrec}{0}
\newcommand{\fixgptfourpctJJJrbtree}{0}
\newcommand{\fixgptfourpctJJJtoml}{21}
\newcommand{\fixgptfourpctJJJpyevtx}{2}
\newcommand{\fixgptfourpctJJJTOTAL}{5}

\newcommand{\autogptthreeJJJcolorsys}{7}
\newcommand{\autogptthreeJJJheapq}{19}
\newcommand{\autogptthreeJJJhtml}{31}
\newcommand{\autogptthreeJJJmathgen}{62}
\newcommand{\autogptthreeJJJstrsim}{50}
\newcommand{\autogptthreeJJJbsdrec}{20}
\newcommand{\autogptthreeJJJrbtree}{26}
\newcommand{\autogptthreeJJJtoml}{33}
\newcommand{\autogptthreeJJJpyevtx}{146}
\newcommand{\autogptthreeJJJTOTAL}{394}

\newcommand{\autogptthreepctJJJcolorsys}{78}
\newcommand{\autogptthreepctJJJheapq}{79}
\newcommand{\autogptthreepctJJJhtml}{74}
\newcommand{\autogptthreepctJJJmathgen}{76}
\newcommand{\autogptthreepctJJJstrsim}{100}
\newcommand{\autogptthreepctJJJbsdrec}{95}
\newcommand{\autogptthreepctJJJrbtree}{96}
\newcommand{\autogptthreepctJJJtoml}{70}
\newcommand{\autogptthreepctJJJpyevtx}{89}
\newcommand{\autogptthreepctJJJTOTAL}{85}

\newcommand{\fixgptthreeJJJcolorsys}{2}
\newcommand{\fixgptthreeJJJheapq}{5}
\newcommand{\fixgptthreeJJJhtml}{11}
\newcommand{\fixgptthreeJJJmathgen}{20}
\newcommand{\fixgptthreeJJJstrsim}{0}
\newcommand{\fixgptthreeJJJbsdrec}{1}
\newcommand{\fixgptthreeJJJrbtree}{1}
\newcommand{\fixgptthreeJJJtoml}{14}
\newcommand{\fixgptthreeJJJpyevtx}{18}
\newcommand{\fixgptthreeJJJTOTAL}{72}

\newcommand{\fixgptthreepctJJJcolorsys}{22}
\newcommand{\fixgptthreepctJJJheapq}{21}
\newcommand{\fixgptthreepctJJJhtml}{26}
\newcommand{\fixgptthreepctJJJmathgen}{24}
\newcommand{\fixgptthreepctJJJstrsim}{0}
\newcommand{\fixgptthreepctJJJbsdrec}{5}
\newcommand{\fixgptthreepctJJJrbtree}{4}
\newcommand{\fixgptthreepctJJJtoml}{30}
\newcommand{\fixgptthreepctJJJpyevtx}{11}
\newcommand{\fixgptthreepctJJJTOTAL}{15}

\newcommand{\fixgptfourbaseJJJcolorsys}{3}
\newcommand{\fixgptfourbaseJJJheapq}{5}
\newcommand{\fixgptfourbaseJJJhtml}{15}
\newcommand{\fixgptfourbaseJJJmathgen}{25}
\newcommand{\fixgptfourbaseJJJstrsim}{5}
\newcommand{\fixgptfourbaseJJJbsdrec}{1}
\newcommand{\fixgptfourbaseJJJrbtree}{9}
\newcommand{\fixgptfourbaseJJJtoml}{15+}
\newcommand{\fixgptfourbaseJJJpyevtx}{15+}
\newcommand{\fixgptfourbaseJJJTOTAL}{93+}
\newcommand{\fixgptfourbaseJJJTOTALTRIM}{93}

\newcommand{\fixtranscryptJJJcolorsys}{0}
\newcommand{\fixtranscryptJJJheapq}{5+}
\newcommand{\fixtranscryptJJJhtml}{NA}    %
\newcommand{\fixtranscryptJJJmathgen}{NA} %
\newcommand{\fixtranscryptJJJstrsim}{NA}  %
\newcommand{\fixtranscryptJJJbsdrec}{0}
\newcommand{\fixtranscryptJJJrbtree}{1}
\newcommand{\fixtranscryptJJJtoml}{NA}    %
\newcommand{\fixtranscryptJJJpyevtx}{NA}  %
\newcommand{\fixtranscryptJJJTOTAL}{NA}   %

\newcommand{\fixskelgptfourJJJcolorsys}{\fixgptfourJJJcolorsys}
\newcommand{\fixskelgptfourJJJheapq}{\fixgptfourJJJheapq}
\newcommand{\fixskelgptfourJJJhtml}{\fixgptfourJJJhtml}
\newcommand{\fixskelgptfourJJJmathgen}{\fixgptfourJJJmathgen}
\newcommand{\fixskelgptfourJJJstrsim}{\fixgptfourJJJstrsim}
\newcommand{\fixskelgptfourJJJbsdrec}{\fixgptfourJJJbsdrec}
\newcommand{\fixskelgptfourJJJrbtree}{\fixgptfourJJJrbtree}
\newcommand{\fixskelgptfourJJJtoml}{\fixgptfourJJJtoml}
\newcommand{\fixskelgptfourJJJpyevtx}{\fixgptfourJJJpyevtx}
\newcommand{\fixskelgptfourJJJTOTAL}{\fixgptfourJJJTOTAL}

\newcommand{\fixskelgptthreeJJJcolorsys}{\fixgptthreeJJJcolorsys}
\newcommand{\fixskelgptthreeJJJheapq}{\fixgptthreeJJJheapq}
\newcommand{\fixskelgptthreeJJJhtml}{\fixgptthreeJJJhtml}
\newcommand{\fixskelgptthreeJJJmathgen}{\fixgptthreeJJJmathgen}
\newcommand{\fixskelgptthreeJJJstrsim}{\fixgptthreeJJJstrsim}
\newcommand{\fixskelgptthreeJJJbsdrec}{\fixgptthreeJJJbsdrec}
\newcommand{\fixskelgptthreeJJJrbtree}{\fixgptthreeJJJrbtree}
\newcommand{\fixskelgptthreeJJJtoml}{\fixgptthreeJJJtoml}
\newcommand{\fixskelgptthreeJJJpyevtx}{\fixgptthreeJJJpyevtx}
\newcommand{\fixskelgptthreeJJJTOTAL}{\fixgptthreeJJJTOTAL}

\newcommand{\skelbaseJJJcolorsys}{3}
\newcommand{\skelbaseJJJheapq}{9}
\newcommand{\skelbaseJJJhtml}{15}
\newcommand{\skelbaseJJJmathgen}{18}
\newcommand{\skelbaseJJJstrsim}{2}
\newcommand{\skelbaseJJJbsdrec}{0}
\newcommand{\skelbaseJJJrbtree}{4}
\newcommand{\skelbaseJJJtoml}{12}
\newcommand{\skelbaseJJJpyevtx}{26}
\newcommand{\skelbaseJJJTOTAL}{89}

\newcommand{\skelspecJJJcolorsys}{3}
\newcommand{\skelspecJJJheapq}{4}
\newcommand{\skelspecJJJhtml}{13}
\newcommand{\skelspecJJJmathgen}{13}
\newcommand{\skelspecJJJstrsim}{0}
\newcommand{\skelspecJJJbsdrec}{0}
\newcommand{\skelspecJJJrbtree}{2}
\newcommand{\skelspecJJJtoml}{13}
\newcommand{\skelspecJJJpyevtx}{13}
\newcommand{\skelspecJJJTOTAL}{61}

\newcommand{\skelchkrfnJJJcolorsys}{\fixgptfourJJJcolorsys}
\newcommand{\skelchkrfnJJJheapq}{\fixgptfourJJJheapq}
\newcommand{\skelchkrfnJJJhtml}{\fixgptfourJJJhtml}
\newcommand{\skelchkrfnJJJmathgen}{\fixgptfourJJJmathgen}
\newcommand{\skelchkrfnJJJstrsim}{\fixgptfourJJJstrsim}
\newcommand{\skelchkrfnJJJbsdrec}{\fixgptfourJJJbsdrec}
\newcommand{\skelchkrfnJJJrbtree}{\fixgptfourJJJrbtree}
\newcommand{\skelchkrfnJJJtoml}{\fixgptfourJJJtoml}
\newcommand{\skelchkrfnJJJpyevtx}{\fixgptfourJJJpyevtx}
\newcommand{\skelchkrfnJJJTOTAL}{\fixgptfourJJJTOTAL}

%% file: sections/0abstract.tex
\begin{abstract}
Translating software between programming languages is a challenging task, for which automated techniques have been elusive and hard to scale up to larger programs.
A key difficulty in cross-language translation is that one has to re-express the intended behavior of the source program into idiomatic constructs of a different target language. This task needs abstracting away from the source language-specific details, while keeping the overall functionality the same.
In this work, we propose a novel and systematic approach for making such translation amenable to automation based on a framework we call {\em program skeletons}.
A program skeleton retains the high-level structure of the source program by abstracting away and effectively summarizing lower-level concrete code fragments, which can be mechanically translated to the target programming language.
A skeleton, by design, permits many different ways of filling in the concrete implementation for fragments, which can work in conjunction with existing data-driven code synthesizers. 
Most importantly, skeletons can conceptually enable \textit{sound} decomposition, i.e., if each individual fragment is correctly translated, taken together with the mechanically translated skeleton, the final translated program is deemed to be correct as a whole. We present a prototype system called \name embodying the idea of skeleton-based translation from Python to JavaScript. Our results show promising scalability compared to prior works. For 9 real-world \py programs, some with more than about $1k$ lines of code, \percentskelauto\% of their code fragments can be automatically translated, while about $5\%$ require manual effort. All the final translations are correct with respect to whole-program test suites.

\end{abstract}

%% file: sections/new-intro.tex
\section{Introduction}
\label{sec:intro}

Automated code translation asks to translate source code written in one programming language to another. 
The task of moving legacy codebases to newer languages and platforms naturally arises in many settings~\cite{msjump,linkcobol}.
Old languages and library dependencies become obsolete and cease to be actively maintained, 
often resulting in an urgent need to move to a more popular platform with well-maintained libraries. 
Code migration is also useful for filling the gaps between software ecosystems by porting popular libraries to languages where similar functionalities are missing.

Despite its importance, satisfactory automated translation has been a long-standing challenge,
even across similar languages~\cite{terekhov2000realities}.
While simple in theory, the naive solution of constructing a compiler
across a pair of languages is often bogus---the quality of output of such a naive translator is not human-readable or maintainable.
Useful and realistic solutions to automated code migration must
meet three key basic requirements. 
First, such a solution cannot compromise correctness---the translated code should be behaviorally equivalent to the source program.
Second, the solution should scale to the size of real-world programs.
Third, the \tododel{space of }translations produced must be readable
and comprehensible by humans, in turn, paving the way for ease of maintenance.
Likewise, the solution must adhere to typical or \emph{idiomatic} programming style in the target language, for example, make reasonable use of APIs and  libraries in the target language;
solutions that compromise on idiomacy are likely to
defeat the purpose of migrating away from the source language. 

Data-driven approaches such as modern large language models (LLMs)
have shown promise towards the third requirement of
idiomacy~\cite{unsupervised,weisz2021perfection,roziere2021leveraging,flourine,ibrahimzada2024repository,yang2024vert}.
Nevertheless, LLM-based code translation techniques struggle to achieve the first two requirements of correctness and 
scale~\cite{icse2024}, and are largely limited
to small benchmark programs that are dozens of lines in size~\cite{icse2024,transmap,pan2023understanding}. 
As the size of programs scales up, the compound effect of 
multiple mistakes across interrelated functions are known
to make the final translation difficult to fix.

This work focuses on \emph{scaling up} automated code translation.
We start with the observation that a practical method that tackles this 
challenge can be obtained by systematically decomposing the 
core task into smaller and simpler sub-tasks, say by breaking the source code into smaller fragments.
Such a scheme naturally allows for the possibility
of leveraging existing techniques, such as code-LLMs, that are well-engineered
to work with translation tasks of smaller scale.
Of course, a straightforward break-and-translate-independently approach is unlikely to work;
the translated smaller fragments may not gel well together and lead to incorrect code, even though each translated fragment may be correct in isolation.

The central contribution of this paper is a framework called \emph{program skeleton},
that allows for scalable and modular decomposition of the translation task.  A program skeleton captures the part of the source program that can be mechanically translated into the target language. It abstracts away the remaining source program implementation details and replaces them with several placeholders in the target language, which can then be concretized with an implementation separately. The end goal is that the final target program is correct, i.e., passes a given set of whole-program tests.

Ideally, program skeletons should have two salient aspects.
First, program skeletons should enable \emph{sound decomposition}, in that, any correct concretization of the placeholders 
in the skeleton is guaranteed to result in an overall correct program.
Second, skeleton code without placeholders can be automatically translated from the source to the target language. As such, the skeleton must be aware of the similarities and differences between the source and the target programming languages.

In this work, we demonstrate the effectiveness of skeleton-based decomposition
in our tool \name, designed to translate code from \py to \js, two of the most popular languages today.
\name generates skeletons by reasoning at a common semantic model of the two languages,
retaining only those parts that have a direct correspondence between the two languages and abstracting away remaining details. For each given \py program and its associated tests, \name analyzes its execution and replaces the elided source fragments with placeholders. Each placeholder carries with it a \emph{local semantic requirement} that the concrete implementation is expected to satisfy.
After translating the skeleton thus generated, fully mechanically, to our target language \js, \name can work with 
existing code synthesizers, including those based on large language models,  directly to find \js implementations for each placeholder separately. 
Any errors caused by unsound synthesizers can be locally corrected for individual placeholder implementations so that they satisfy the local semantic requirements.

We use program tests both in generating skeletons and checking their correctness. While one can consider formal specifications to capture correctness, the source and target language we consider are dynamic scripting languages for which cross-language functional equivalence checking is notoriously difficult. We have, therefore, chosen to demonstrate the concept of skeletons with a purely test-driven approach for pragmatic reasons.

We demonstrate the performance of \name on \py to \js translation using a benchmark considered in recent work~\cite{transmap}, which considers programs larger than prior works~\cite{duoglot}, and we extended it further to include programs nearing $2k$ lines of code.  $4$ out of $9$ of the translated programs can directly pass whole-program tests without any human intervention. The remaining require a few code fragments to be manually fixed, owing to the limitation of the off-the-shelf LLMs we employed for placeholder synthesis. A total of $\percentskelauto\%$ of the code fragments require no manual intervention and are translated to code that is correct with respect to test cases. After the remaining $5\%$ are manually fixed, all translated programs pass the given test cases. We thus conclude that \name offers a promising avenue towards mostly automated translation for \py to \js.

%% file: sections/2problemoverview.tex
\section{Program Skeletons: A Preview}
\label{ref:problemoverview}

\begin{figure}[t]
    \centering
    \includegraphics[trim=0.2in 0.9in 0.2in 0.7in, clip, width=0.97\linewidth]{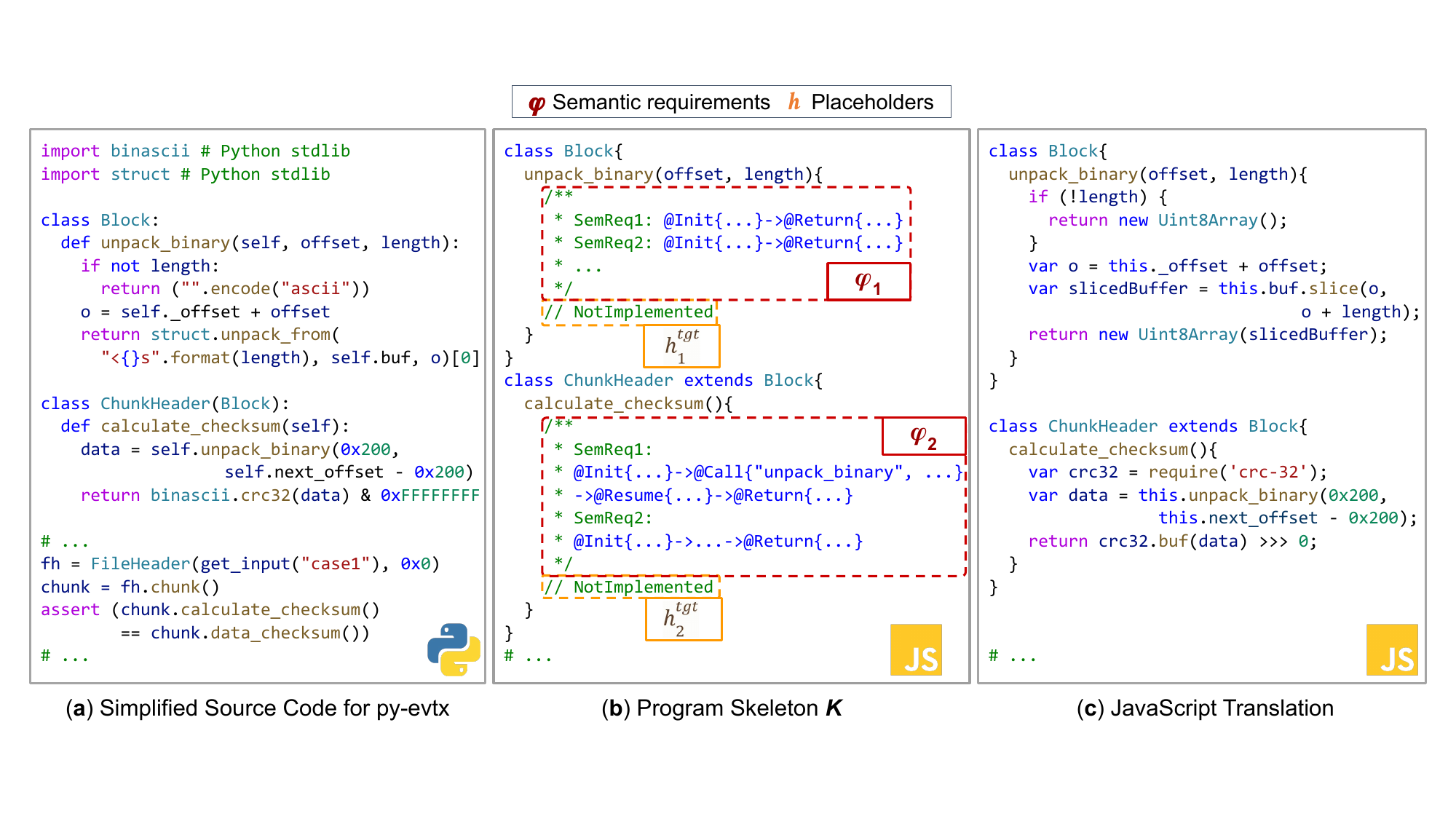}
    \caption{\todoadd{An illustrative program skeleton:} (a) The simplified source code for one program named "py-evtx" (2k lines of code) in our benchmark, (b) the program skeleton\tododel{ generated by \name} for the translation, and (c) the complete translation\tododel{ produced by \name} obtained by filling the code fragments into the skeleton.}
    \label{fig:moexp}
\vspace{-10pt} 
\end{figure}

We begin by concretely illustrating the concept of a skeleton using an example. The \py source program shown in Fig. \ref{fig:moexp}(a) is to be translated into \js. 
Fig.~\ref{fig:moexp}(b) shows the program skeleton generated for the \py program, but translated to adhere to the syntax of \js language. The skeleton is a partial \js program with {\em placeholders} along with local semantic requirements for each placeholder specified. The placeholder replaces some of the detailed implementation that was present in the \py code. The skeleton can be completed subsequently by an independent {\em fragment synthesis} step, which generates concrete \js code fragments that will replace the placeholder in the skeleton eventually. The fragments filling a placeholder are expected to satisfy its local semantic requirement. 
The end result after such fragment synthesis is a runnable \js program shown in Fig.~\ref{fig:moexp}(c).

The program skeleton approach thus gives us a clean separation of concerns between two parts of the translation process: skeleton generation and skeleton completion. The generated skeleton code, as shown in Fig.~\ref{fig:moexp}(b), is mechanically generated using a rule-based system. The final completion can make use of any off-the-shelf synthesizer for fragments, including LLMs.

In the following, we formalize the notion of a skeleton and its ideal properties. A skeleton can be seen as an intermediate representation that conceptually has two parts: a syntactic representation and annotations for semantic requirements of individual placeholders.

\myparagraph{Program Skeleton: Syntactic}{
A syntactic skeleton is a program with ``holes'' or ``placeholders''.
Formally, a syntactic program skeleton for a language $\lang$ is a partial program $\skl$ with holes $\holes{\skl} = \set{h_1, \ldots, h_n}$.
A {\emph completion} of skeleton $\skl$ is simply a mapping $\cmpl : \holes{\skl} \to \lang$; we will use $\cmpl(\skl)$ to denote the complete program induced by the completion $\cmpl$.
}

\myparagraph{Program Skeleton: Annotated}{
Naturally, not all completions are expected to be desirable. 
A natural way to constrain possible completions is to specify independent requirements for each placeholder.
We annotate each placeholder $h$ in skeleton $\skl$ with a \emph{local semantic requirement} $\lsemreq \subseteq \lang$\ignore{ (also closed w.r.t. $\eq$)}.
Formally, a completion $\cmpl$ is said to be consistent with local semantic requirements $\set{\lsemreq_i}_i$ if
$\cmpl(h_i) \in \lsemreq_i$ for each placeholder $h_i$.
We say that $\sklsem = \langle \skl, \lsemreq_1, \lsemreq_2, \ldots, \lsemreq_n \rangle$ is an \emph{annotated program skeleton}, and will often abuse the term program skeleton for it.
}

A skeleton should ideally enable a {\em sound decomposition}. Informally, this means that if each placeholder is correctly synthesized, the final completed skeleton yields a correctly translated program. We now formalize this goal using an abstract notion of correctness between source and target programs, $\pgm_\src$ and $\pgm_\tgt$, respectively. Note that whether a translation is considered correct depends on what programs in the target language are considered equivalent to a given source program. We assume the existence of such a notion of behavioral equivalence between two programs across languages and encode it as a binary relation $\eqsrctgt \subseteq \lang_\src \times \lang_\tgt$. 

\myparagraph{Sound Decomposition} 
Let $\eqsrctgt$ be a relation capturing some notion of equivalence between source and target language programs. We say a skeleton $\hat{\skl}_\tgt$ is \emph{sound} with respect to $\eqsrctgt$ and a program $\pgm_\src \in \lang_\src$, if for every completion $\cmpl_\tgt$ of $\hat{\skl}_\tgt$, 
we have $(\pgm_\src, \cmpl_\tgt(\skl_\tgt)) \in \eqsrctgt$.

\vspace{0.1in}
\noindent
{\em Remark.} The above notion of soundness is an abstract one---it does not give concrete definitions for $\eqsrctgt$, which defines program equivalence or correctness. Throughout this work, the specific notion of equivalence we use is an empirical one, and asks that the source and target programs behave the same on a set of user-defined tests. The final solution we present uses soundness as a guiding principle, and our implementation is a best-effort illustration of the concept of skeletons. In particular, we provide neither a complete specification of cross-language equivalence relation, nor do we claim that our implementation guarantees soundness generically for all programs and all possible program inputs.

Our system \name aims to be a practical prototype. To do so, \name defines an intermediate \tododel{language}\todoadd{representation} called \decocr which captures the commonality of the source \py and target \js language and is amenable to mechanical rule-driven translation. The main practical challenge in designing this representation is to obtain the right level of expressiveness. An overly restrictive intermediate representation may make it difficult for off-the-shelf synthesizers to find any idiomatic implementation for the placeholders. On the other hand, a skeleton that allows for a larger set of possible completions gives more freedom on which implementations to choose (i.e., synthesize), giving room for idiomatic translations, and thus has an overall better utility. 
We empirically demonstrate this utility of \name in our evaluation (Section \ref{sec:eval}), where we show that \name can successfully translate real-world benchmark programs mostly automatically.

\myparagraph{Example Revisited} 
The skeleton represented in our \decocr carries the high-level syntactic structure of the target program, and includes lexical scopes and function signatures. These abstractions are indeed similar across both the source and target languages, and for this reason, can be mechanically translated using a fixed set of rules.
For example, Fig.~\ref{fig:moexp}(b) shows the \js program skeleton derived from the \py program in Fig.~\ref{fig:moexp}(a). Observe that, the lexical scoping and nesting structure of function and class definitions have been preserved across the \py program (Fig.~\ref{fig:moexp}(a)) and the translated skeleton (Fig.~\ref{fig:moexp}(b)).

The skeleton at this stage is an incomplete \js program with placeholders (dashed rectangles), each of which carries a local semantic requirement. In our setting, these requirements are encoded as I/O sequences that specify the \emph{observable effects} (details in Section \ref{sec:decooverview}) that the code fragment implementing the placeholder must produce.
Fig. \ref{fig:moexp}(b) highlights the semantic requirements for the placeholder marked as $h_2^\tgt$ as an example.
Informally, the illustrated semantic requirements shown in the figure can be understood as --- \emph{when executing the code fragment from a certain initial state (\code{@Init \{...\}}), the code fragment should first call a function named \code{unpack_binary} with specific arguments (\code{@Call \{...\}}), and then after obtaining the result of the call, finish its remaining computation and return a specific value (\code{@Return \{...\}}).}
Such semantic requirements mirror how the corresponding fragments must interact with the skeleton as well as with each other.

The compositional nature of skeleton-based translation is evident from the example. The skeleton in Fig. \ref{fig:moexp}(c) shows a completion for placeholders in the skeleton as generated by a modern LLM. All of the different ways of implementing $h_2^\tgt$, as long as valid according to $\lsemreq_2^\tgt$, should compose well with completions that are valid for other placeholders such as $h_1^\tgt$. 
In other words, if the individual placeholder fragments are correct as per the local semantic requirements (encoded as annotations $\varphi_i^{\tgt}$), then the completion of the whole translation should automatically satisfy the global semantic requirement, which in our work corresponds to passing the tests.

An important consequence of this compositional nature is that one can locally check if a candidate code fragment has the expected behavior as determined by the requirement of its placeholder. 
That is, errors in the completion of one placeholder get caught locally rather than affecting the semantics of other placeholders in unpredictable ways.
This ability to isolate and localize errors in the completion allows us to leverage expressive but unsound synthesizers, which is typical of modern data-driven approaches.

\color{black}

%% file: sections/3decooverview.tex
\section{\name: Overview of Skeleton Generation \& Completion}
\label{sec:decooverview}

In this section, we outline our key design choices for skeleton generation and skeleton completion. At a high level, the skeleton generation is guided by the high-level similarity of the two languages. Section \ref{sec:overviewsyn} outlines how we leverage this similarity to determine the syntactic skeleton from the source program. In Section \ref{sec:overviewsem}, we discuss how we obtain the local semantic requirements for placeholders by, in turn, modeling and distilling the observable behaviors of each fragment from the source. 
Finally, in Section \ref{sec:overviewsynth}, we discuss our practical solution to obtain the complete translation.

\subsection{Determining the Syntactic Structure of the Skeleton}
\label{sec:overviewsyn}

\py and \js are similar at a high level 
but dissimilar at lower levels.
There are many similar aspects in their high-level design. 
For example, they offer similar control-flow constructs, such as loops and if-conditions. 
They also have similar lexical scoping rules and closures for capturing non-local variables. 
Both languages are dynamically typed and have many commonly used data types that have similar semantics. For example, \code{List} and \code{Dict} in \py are similar to \code{Array} and \code{Object} in \js, respectively.
Both languages support object-oriented encapsulation by allowing class definitions with associated methods.

However, statement-level details of program representation can substantially differ across the two languages. The most obvious differences are in the available standard library APIs and their semantics, which force idiomatic translation to express the source program logic using a different set of APIs and operators in the target program. 

\name generates a program skeleton that unifies the high-level syntactic structure between the source and the translation, i.e., lexical or function scopes, along with the symbol table for each scope.
\name assumes that such function- or class-level structure is thus fully specified by the source. We expect the user to adjust the source structure before using \name if a different high-level structure is preferred for the translation.
Such a unification of high-level program structure between source and intended translation can simplify the analysis of local semantic requirements for placeholders later on, since
\name can see the source program as a ``completed skeleton'' $\Gamma_\src(\skl_\src)$ but in the source language,
which yields a mapping between placeholders in the skeleton ($h^\src_0, ..., h^\src_n$) to code fragments in the source program ($\frag^\src_0, ..., \frag^\src_n$). 
We will explain how we determine the semantic requirements for placeholders in the next sub-section.

\ignore{The major challenge is to come up with local semantic requirements for each placeholder. due to dissimilarities between languages that affect implementation details.}

\subsection{Observable Effects for Semantic Requirements}
\label{sec:overviewsem}

At a high level, our semantic requirements are extracted from the original program in the form of input-output behaviors for each of the code fragments. Of course, care must be taken to determine the level of detail that must be captured in such I/O behaviors; ideally, we would like to capture the essential details to ensure the soundness of the skeleton, while removing unnecessary details from the semantic requirement to allow idiomatic implementation in the target language.

The challenge in capturing precisely the semantics of realistic \py and \js programs is that they have ``messy'' behaviors: the semantics of code fragments are far more complicated than pure functions on primitive values.
What should be considered as inputs and outputs soon becomes unclear in the presence of closures, shared data references, and their interactions with numerous, potentially higher-order, library APIs.

A naive solution to precisely determine the input-output behaviors of a code fragment may capture all the state changes in the underlying language runtime.
This will not miss any details but is not useful for our skeleton generation task because the captured state changes can involve many language-specific details. Examples of such low-level details include internal implementations of iterators, library APIs, temporary closures, and special control-flow states; these cannot be easily translated into idiomatic constructs in the target language.
On the other hand, a coarse-grained analysis that leaves out low-level details comes at the risk of creating errors in the resulting decomposition, i.e., code fragments satisfying their coarse-grained semantic requirements may not result in a runnable and correct target program when merged back into the skeleton.

\ignore{For each placeholder $h_i$, we mirror the behavior of the corresponding source code fragment $\frag^\src_i$.}

To address this challenge of determining the right level of abstraction at which we must track the input-output behaviors, \name takes guidance from the following key design principle:

\begin{description}
\item[\sc Indistinguishability Principle:] {\em Any two code fragments that satisfy the same semantic requirements of a placeholder should not be distinguishable from outside of the placeholder. }
\end{description}

This principle captures whether the synthesized fragments for a placeholder can be composed correctly with the rest of the program. Specifically, 
instead of asking whether the language interpreter can differentiate between completions of this placeholder,
the principle outlined above
asks whether the rest of the code fragments (besides the one in consideration) can tell the difference in the observed behavior of the code fragment in consideration. We use the phrase \emph{observable effects} to describe such externally observable behaviors. The indistinguishability principle is a pragmatic relaxation that allows for greater flexibility in skeleton completion than the aforementioned naive solution. Specifying code fragments based on observable effects does not require full state equivalence between source and target programs. %

\myparagraph{The Need for a Common Model}
It is not straightforward to accurately extract observable effects for code fragments. There is no standard way to differentiate ``internal'' v/s ``externally observable'' effects---it depends on how we model program semantics and whether such modeling can be mapped to both the source and the target language.
In response, we first determine a common model of program semantics, called \procint, which makes observable effects explicit.
The salient aspect of the common \procint model is --- any two programs with the exact same semantics must have indistinguishable code fragments (same observable effects).
\name maps concrete semantics in both \py and \js to this common model; intuitively, when two programs $\pgm_\src$ and $\pgm_\tgt$ in these languages that get mapped to semantically equivalent programs in \procint, they can be considered equivalent. We describe \procint at a high level next, while deferring details to Section \ref{sec:design}.

\myparagraph{The \procintbold Model} We propose a common model named \procint that treats code fragment executions as standalone ``processes''. 
Execution of a program becomes a collection of "communicating processes", instead of a single process in the real language interpreter. 
Processes are isolated, and they can only communicate through messages, similar to the typical concept of processes in process algebra~\cite{milner1980calculus} 
except for the difference that there is no parallel execution.
Interactions between code fragments can be mapped to several types of ``communication messages''.

A (\py or \js) program, when abstracted in the \procint model, ``emulates'' each invocation of every code fragment $g_i$ (filling placeholders $h_i$) in a separate stateful ``process''. 
In this sense, \procint  ``executes'' a full program $\pgm$ with skeleton $K$ as the following communication sequence between ``processes'':
$\rho = \procint(\pgm)  = \mathcal{K}  \myarrow{{Init}_1} \textcolor{violet}{P_0} \myarrow{{Call}_2} \mathcal{K} \myarrow{{Init}_3} \textcolor{violet}{P_1} \cdots \myarrow{{Res}_{m-1}} \textcolor{violet}{P_0} \myarrow{{Ret}_{m}} \mathcal{K}$. 
Here, each $P_i$ is a process that corresponds to some invocation of a placeholder, and $\mathcal{K}$ is the unique process corresponding to the skeleton $K$. A process can be interrupted and resumed multiple times during an execution sequence $\rho$; these transitions correspond to control flow transfers, and they are denoted with arrows (such as $\myarrow{{Res}_{m-1}}$) in $\rho$. Each control flow transfer is accompanied by the exchange of data, which is captured with ``messages''. For example, in the sequence $\rho$ here, process $\mathcal{K}$ first communicates to the process $\textcolor{violet}{P_0}$ with message ${\sf {Init}_1}$, and in response, 
$\textcolor{violet}{P_0}$ sends a message ${\sf {Call}_2}$ to process $\mathcal{K}$.

\myparagraph{\todoadd{Determining} Observable Effects}
\todoadd{With the \procint model in place, the problem of obtaining observable effects reduces to extracting the communication sequence $\rho$ from concrete program executions. While the process boundaries are clarified by the skeleton structure, determining what information to include in the inter-process messages requires careful thought. The goal here is to keep the communicated information (1) minimal yet (2) sufficient. Ideally, we must include `just enough' information so that (sufficiency) the concrete execution of the involved code fragment can be correctly ``emulated''  in an isolated process , and (minimality) removing any more information can cause such an emulation to either completely fail or diverge from the real execution.
The details of our \procint are designed to align closely with this goal, and may vary depending upon the languages in question (\py and \js in our case). We provide these details in Section \ref{sec:design}.}

\myparagraph{Ideal vs. Implementation}
\todoadd{We remark that the \procint model directly lends itself to a precise notion of equivalence for programs across languages, paving the way for sound decomposition. }
More precisely, the (ideal) observational equivalence  induced by \procint is the relation $\eqsrctgt^{\text{Ideal}} = \{(\pgm_\src, \pgm_\tgt) ~|~ \mathcal{F}(\procint(\pgm_\src)) = \procint(\pgm_\tgt), \pgm_\src \in \lang_\src, \pgm_\tgt \in \lang_\tgt\}$, where the skeleton of $\pgm_\src$ (and $\pgm_\tgt$) used by \procint is the unique skeleton constructed by \name. Here, $\mathcal{F}$ is a cross-language type mapping to `retrofit' objects in the source language to those in the target language; we present more details in Section~\ref{sec:semstep2}.
In principle, the `ideal' \name translates a given program and succeeds if the ideal check `$\mathcal{F}(\procint(\pgm_\src)) = \procint(\pgm_\tgt)$' succeeds.
\todoadd{Our implementation of \name is, however, best-effort. In practice, our equivalence checks are only approximate due to various practical concerns ---  floating point errors, imprecise modeling of APIs, and time overhead when comparing large objects, to name a few. 
As a result, our implementation deviates from the ideal and may therefore not guarantee soundness in the absolute sense. 
At the same time, we remark that our empirical evaluation demonstrates that this deviation is not high, as is evident by the fact that, whenever an error in translation is observed, it can always be pinpointed to the code fragment where the error originates.}

\subsection{Code Fragment Synthesis}
\label{sec:overviewsynth}

Once we have the semantic requirements for the skeleton (obtained by executing the source and then extracting the sequence $\rho$ in \procint), one can already complete each placeholder $h_i$ naively --- compile all messages involving the process instances of \todoadd{the code fragment} $g_i$ \todoadd{(to be synthesized)} into a single semantic requirement $\lsemreq_i$ and subsequently query the synthesis engine with the specification $\lsemreq_i$.
Unfortunately, this approach overwhelms the synthesizer since such a naively encoded specification can be very large, containing many rounds of messages arising from multiple process instances of the same fragment. In such a setting, a data-driven synthesizer (read: LLM) is likely to make many mistakes that compound and are often hard to debug.

In \name, we instead opt for a different approach --- gradual refinement with spot-checking.
Here, the code fragment $g_i$ corresponding to some placeholder $h_i$ is incrementally refined using an increasingly more thorough sequence of specifications $\varphi_i^{(1)}, \varphi_i^{(2)}, \cdots, \varphi_i^{(j)}$ for it.
Operationally, this is performed by traversing the code fragments in the order induced by the sequence $\rho$. At each step $m$ of the sequence, the placeholder (say $h_i$) at step $m$ is considered for synthesis. If this is the first time $h_i$ is encountered,  we \tododel{do the usual --- }query the synthesizer with the specification $\varphi_i^{(1)} = \psi_m$ and retry until success; here the specification $\psi_m$ encodes the input-output message pair at step $m$. If, on the other hand, we do have an earlier synthesized code $g_i$ for $h_i$ (which satisfies the specification $\varphi_i^{(j)}$ so far), we proceed as follows. If $g_i$ already meets the additional specification $\psi_m$ induced by the new input-output pair at step $m$, we simply proceed to the next step, while keeping the specification $\varphi_i^{(j)}$ as is. Otherwise, we update the specification to $\varphi_i^{(j+1)} = \varphi_i^{(j)} \land \psi_m$ and then query the synthesis engine with this updated specification (\emph{gradual refinement}).
As before, upon success, we move to the next step; however, if the check fails, we immediately retry synthesis for the current fragment $g_i$ (\emph{spot checking}).
This approach offers two key benefits that are instrumental in making \name practical.
First, the gradual refinement step is designed so that it only accounts for new input-output specifications that actually fail during the process (instead of \emph{every} input-output pair), making our specifications small in practice, and is similar in spirit to CEGIS approaches~\cite{sketch08}.
Second, unlike the vanilla `synthesize-everything-and-check-at-the-end'  approach, spot checking pro-actively finds errors induced by the synthesizer (read LLM) and corrects them before proceeding to do more synthesis work for other placeholders.
We present more details in Section~\ref{sec:codesynth}.

%% file: sections/4_1_syntax.tex
\section{Design Details of \name}
\label{sec:design}

\name generates the two main parts of a program skeleton, i.e., syntactic structure and semantic requirements, by analyzing the source program. 
After that, it synthesizes and refines code fragments for placeholders following the execution order.
To explain details of our design, we will use a small standalone \py program shown in Fig.~\ref{fig:ovsyntax} and walk through how it is translated to \js.

\input{figtabs/fig_ov_syntax_sim}

\subsection{Syntactic Structure of the Program Skeleton}
\label{sec:skeletonsyntax}

\name determines the syntactic structure of the skeleton from the source by mirroring the function signatures and the symbols accessible across the lexical scopes, while leaving low-level implementations as placeholders.
\ignore{\umang{Didnt the reviewer explicitly pointed out their discomfort with an ujustified use of the term `mirroring?}}
Referring to our running example, the \py program in Fig.~\ref{fig:ovsyntax} (left) is a typical multi-function program with closures.
The program consists of four lexical scopes, one for global scope and three for different functions. The entry point is the code fragment in the global scope, which also serves as the unit tests for the program.
Closures are created and passed around as values in this program.
The function \code{_update} is a closure within the function \code{compute} and is passed (in line 8) to another function \code{multi}.
An idiomatic and correct \js translation of the \py source is shown on the right in~Fig.~\ref{fig:ovsyntax}.
The example highlights the following similarities and dissimilarities between the \py program and the \js translation:

\begin{description}
    \item[\it Largely Similar: Lexical Scoping.] 
    The source and the translation have highly similar function declarations and nesting structures of lexical function scopes (\highlightlightgreen{green} in Fig. \ref{fig:ovsyntax}), if we omit several non-escaping closures in the code (such as the \code{i => a[i] *= n} in \js). Our program skeleton keeps such lexical scoping information, which can be translated largely as-is to \js.\footnote{Note that the example does not show certain features such as keyword arguments in \py, which has no direct correspondence in \js. Details on how we address them are presented later in Section \ref{sec:impl}.}
   
    \item[\it Partly Similar: Symbol Tables.] 
    The source and the translation have similar but not identical symbol tables. Some symbols in the source are retained in the translation, while others are eliminated. For example, the \code{asum} variable is kept while the \code{x} variable in the \code{compute} function is eliminated and replaced by an expression \code{arr[i]}. A similar elimination happens for the variable \code{y} in \code{multi}. The common characteristic between eliminated variables is that they are not accessed outside their scope. \name statically eliminates such symbols and produces the skeleton without them.
   
    \item[\it Dissimilar: APIs usage and Coding Conventions.] 
    While two languages share similar high-level structures, their individual statements often differ. Semantically similar statements are typically expressed in different APIs, operators, and coding styles. These differences further affect how the program logic is structured. 
    For example, \py APIs such as \code{range} and \code{zip} (\highlightorange{orange} in Fig. \ref{fig:ovsyntax}) are often used to write loops. However, there is no natural direct analogue for the \code{range} or \code{zip} API in \js. 
    Idiomatic translations of such loops in \js will likely use different kinds of APIs, such as \code{keys} and \code{forEach}. 
    Owing to such language differences, \name leaves statement-level details as placeholders in the skeleton and reconstructs them semantically in the target language.
\end{description}

\input{figtabs/fig_ov_syntax_tabs}

\input{figtabs/defsyn}

\name aims to produce program skeletons that preserve the similarity between the source and the translation while abstracting away the differences. It thus views the source program as a completion of the source skeleton  $K^\src$ with source fragments $\fragsrc_0, ..., \fragsrc_3$.
A syntactic skeleton of the \js program can then be generated by referring to $K^\src$. Code lines that are part of the resulting skeleton are marked as $K^\tgt$ at the top-right corner of Fig. \ref{fig:ovsyntax}.

The right-hand side of Fig.~\ref{fig:syntax_of_skeleton} shows a graph representation of the syntactic structure of the skeleton in our \decocr. 
It includes four lexical scopes with associated symbol tables while omitting statement-level details.  
The parent-child relation corresponds to the nesting structure of lexical scopes.
Each symbol table lists all declarations in the corresponding lexical scope, including identifiers for variables and nested functions (closures). Generators and classes are conceptually similar to closures. 
Symbols not accessed outside the current scope are eliminated.
The closure \code{_update}, despite being nested in \code{compute}, may escape and thus is kept in the skeleton to allow correct modeling of its semantics.
The syntactic structure of the skeleton is formally expressed in \decocr as in Fig.~\ref{fig:grammarsyn}, which consists of two parts: the observable symbols for each scope ($\textbf{Scopes}$) and the mapping of symbols across scopes ($\textbf{Relations}$).
Each scope has a symbol table. For example, the observable symbols of the four symbol tables in Fig. \ref{fig:syntax_of_skeleton} correspond to the $\textbf{SymTab}$ for each scope ($\textbf{Id}^\text{Scope}$).
Symbols may be related across scopes. For example, the dashed arrows in Fig. \ref{fig:syntax_of_skeleton} show what non-local symbols are referring to, and the green arrows show the parent-child relation of scopes.
These relations correspond to $\textbf{Map}^\text{nonlocal}$ and $\textbf{Map}^\text{Closure}$ in Fig. \ref{fig:grammarsyn}.

The example so far explains common language features in \py and \js that have direct correspondence to \decocr in Fig.~\ref{fig:grammarsyn}. The handling of other language features we supported, such as class declarations and decorators, are further explained in 
\techref{Appendix~\ref{app:norm}}{our technical report~\cite{supp}},
where we explain details of source code normalization.

%% file: figtabs/fig_ov_syntax_sim.tex
\begin{figure}[t]
    \centering
    \includegraphics[trim=1.9in 1in 1.9in 2.0in, clip, width=0.8\linewidth]{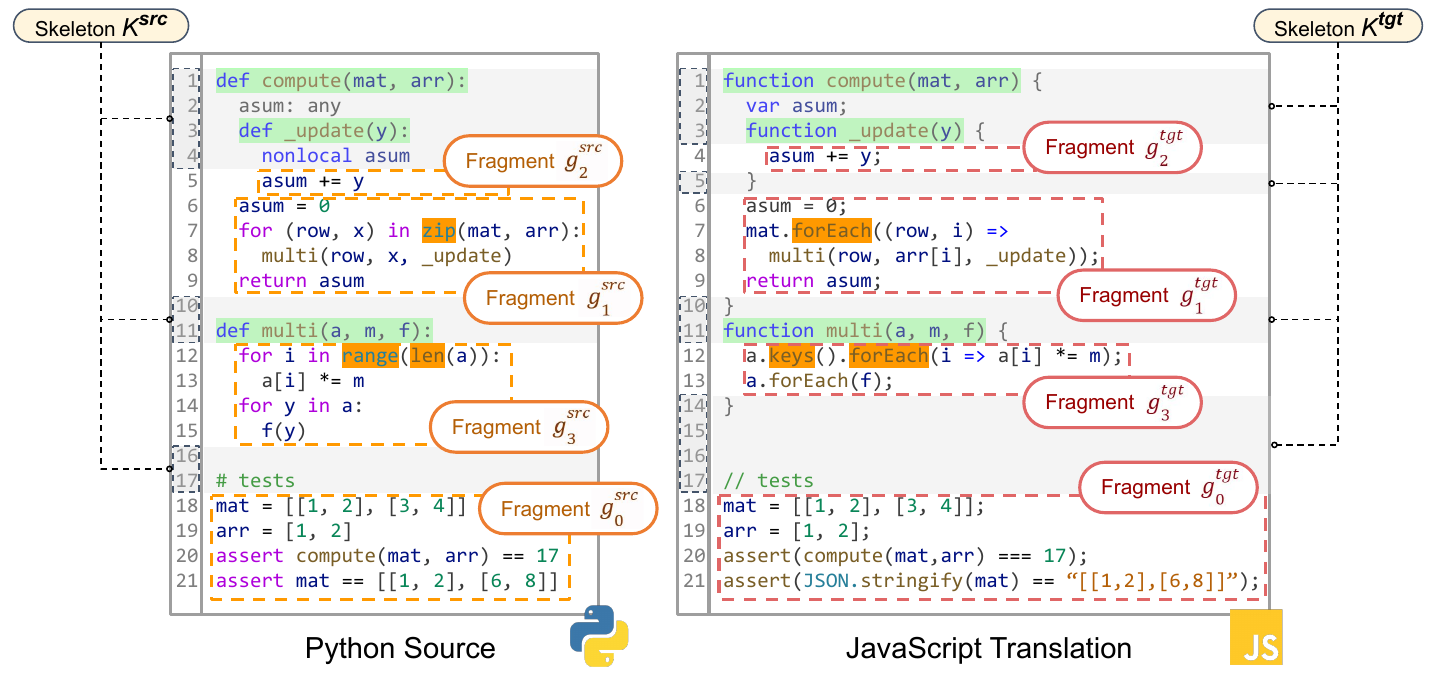}
    \caption{An example Python program (left) and its JavaScript translation (right). The syntactic structures of these two programs are similar at the level of lexical scopes, but different at the level of  individual statements.}

    \label{fig:ovsyntax}
\end{figure}

%% file: figtabs/fig_ov_syntax_tabs.tex
\begin{figure}[ht]
    \centering
    \includegraphics[trim=1.4in 1.4in 1.4in 1.3in, clip, width=0.8\linewidth]{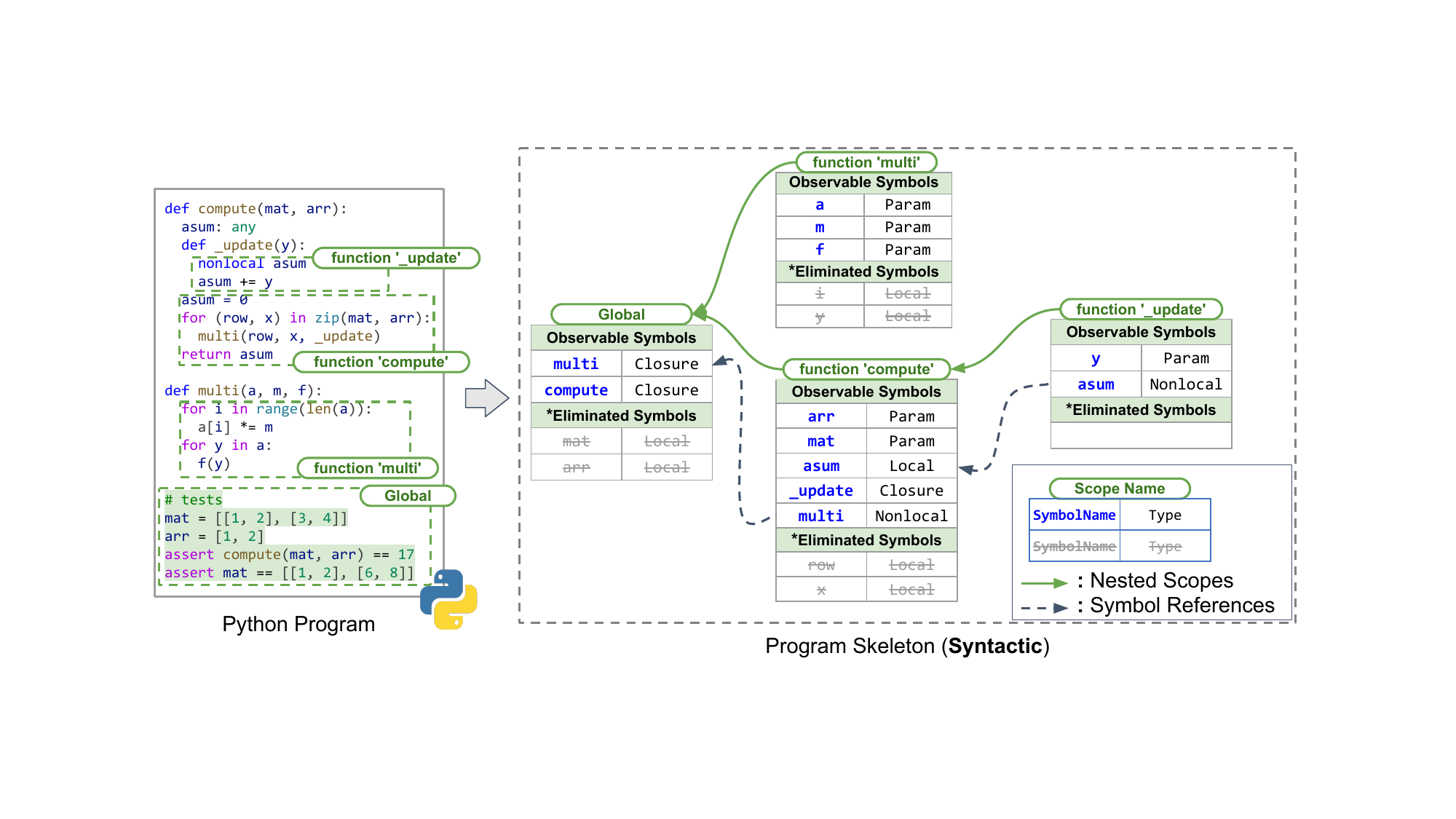}
    \caption{A graph representation of the syntactic structure of the skeleton}
    \label{fig:syntax_of_skeleton}
\end{figure}

%% file: figtabs/defsyn.tex
\begin{figure}[!htbp]
    \centering
    \footnotesize
    \begin{tabular}{@{}r@{\hskip3pt}c@{\hskip3pt}p{4.2cm}p{5cm}@{}}

        \textbf{\decocr} & ::= & $(\textbf{Syntax}, \textbf{Semantics})$ & // The grammar of {\decocr} \\
        \textbf{Syntax} & ::= & $(\textbf{Scopes}, \textbf{Relations})$ & // The syntactic structure of the program \\
        \textbf{Scopes} & ::= & $\textbf{Id}^{\text{Scope}} \to \textbf{SymTab}$ & // The structure is a set of lexical scopes \\
        \textbf{SymTab} & ::= & $(\chi^\text{Param}, \chi^\text{Local}, \chi^\text{Nonlocal}, \chi^\text{Closure})$ & // The observable symbol table \\
        $\chi$ & ::= & $\overrightarrow{\textbf{Id}^{\text{Sym}}}$ & // A list of symbols \\
        $\textbf{Relations}$ & ::= & ($\textbf{Map}^\text{nonlocal}, \textbf{Map}^\text{Closure}$) & // Relations between symbols \\
        $\textbf{Map}^\text{nonlocal}$ & ::= & $\chi^{\text{Nonlocal}} \to \chi^{\text{Captured}}$ & // Mapping of non-local symbols \\
        $\textbf{Map}^\text{Closure}$ & ::= & $\chi^{\text{Closure}} \to \textbf{Id}^{\text{Scope}}$ & // Mapping of closure symbols \\
    \end{tabular}

    \caption{The grammar of \decocr (part of it) that describes the syntactic structure of program skeleton.}
    \label{fig:grammarsyn}
\end{figure}

%% file: sections/4_2_semantics.tex
\subsection{Extraction of Semantic Requirements}
\label{sec:skeletonsemantics}

Our \procint model mentioned earlier in Section \ref{sec:decooverview} gives us a unified representation of the concrete semantics of both the \py source and the \js translation. 
\name constructs the semantic requirements for placeholders in two steps. 
First, we 
record the observable effects of each source fragment under concrete program inputs.
Then, the observable effects are directly mapped into semantic requirements for the corresponding placeholder.

\subsubsection{Obtaining Observable Effects} 
\label{sec:semstep1}

We construct a dynamic analyzer, which is named \procint analyzer, that monitors actual program execution to extract messages corresponding to our \procint execution model, where each invocation of a code fragment is a ``process''. Observable effects for each code fragment will be the relevant messages between ``processes''.

\input{figtabs/figworkflow-stepmodel}

For our running example of Fig. \ref{fig:ovsyntax}, \name models the source \py program as 9 communicating processes visualized in Fig. \ref{fig:workflowstepmodel}.
The whole message sequence $\rho$ is visualized as arrows between processes.
In this semantics model, the "skeleton process" $\mathcal{K}_\tgt$ (created from $K^\tgt$) orchestrates all other 8 "fragment processes", $P_0$, ..., $P_7$. Some code fragments (such as $g_3^\tgt$) are invoked multiple times and thus have multiple corresponding processes (e.g., $P_2$ and $P_5$).

The execution of programs under our specific \procint design can be summarized as follows.
First, there is only one process executing at any point in time. The execution starts from the skeleton process $\mathcal{K}$, which immediately starts the code fragment process corresponding to the entry point (which can be the tests for the whole program).
When a code fragment process transfers control flow into other parts of the program, it pauses itself after sending a message that wakes up the central coordinator---the skeleton process. 
The skeleton process decides the next step of execution, either resuming an existing process (e.g., for return or throw) or creating a new process (e.g., for call).
Communication occurs only during control flow transfers. The communicated messages not only transfer control flow but also contain data needed for later execution.
Details on the semantics of the skeleton (in our \procint execution model) are in
\techref{Appendix~\ref{app:sem}.}{our technical report~\cite{supp}.}

\myparagraph{Behaviors Captured}
To be correct, the analysis should capture a sufficient level of detail to distinguish different executions later on.
The level of detail considered sufficient depends on possible program behaviors in the language. 
\name supports subsets of \py and \js ($\lang_\src$ and $\lang_\tgt$) where all interactions between code fragments can be categorized into 3 kinds of control flow interactions (calls, returns, and exceptions) and 3 kinds of data sharing (data passing, shared variables, and shared references).
\todoadd{Each closure value is modeled as a unique tuple of two Ids: the scope Id of the closure itself and the process Id of its creating process. Such tuples allow our \procint model to correctly ``emulate'' closure invocations.}
Shared references require careful consideration since they can appear in nested data objects, which will be explained next.

\myparagraph{Behaviors Left Out} Our model views certain program behaviors and states as the internals of a ``process'' and are thus not captured in our \procint semantics. Examples of such internals include binding changes to eliminated local variables and invocations of eliminated not-escaping closures (mentioned in Sec. \ref{sec:skeletonsyntax}), as well as exception objects raised and caught within the same fragment.
The tricky part is about shared data references. Variables and objects created in one code fragment may be accessed in other code fragments in many ways, either by escaping closures, shared references on the heap, or higher-order library APIs that pass them to different code fragments.
\todoadd{Our \procint analyzer
dynamically maintain a set of \emph{may}-access objects for each process
to keep track of objects accessible by each of them. We name such object sets as \emph{observable} sets. The observable set for each process is updated at the time of control-flow transfers, where objects reachable from observable symbols or objects will be added to the set and remain observable until the end of the corresponding process.}
Other objects are not included in the observable set, such as most of the temporary objects or internal objects in certain library API implementations.
\ignore{Our may-access analysis aims to over-approximate the set of processes that access the current state of each shared data object.}
The analysis eliminates objects from communication messages when their current state is not accessed by other processes.

\input{figtabs/defsem}

\input{figtabs/fig_ov_semantics}

\myparagraph{Side Effects of APIs} 
The API calls in the program may also contribute to the observable effects of code fragments and are modeled in \procint. As mentioned in section \ref{sec:skeletonsyntax}, many of the APIs do not have clear mappings between \py and \js. Thus, we aim to abstract them away when possible, rather than modeling them as {\sf Call} effects. 
Specifically, we categorize side effects by non-pure API calls into two categories, namely, \emph{transparent} effects and \emph{opaque} effects. API calls result in transparent effects when they mutate or create data objects that can be referred from the inputs or outputs of those APIs. We consider these effects as ``transparent'' since it suffices for the \procint analyzer to track changes in the observable objects to capture their effects. Opaque effects, on the other hand, come from APIs that interact with the external environment or mutate hidden program states. Notable examples are \code{print} and \code{random} APIs in \py. Our \procint analyzer models such opaque effects as special kinds of call messages to the skeleton ``process'', which are handled by the skeleton ``process'' directly. 
In the actual implementation, we model such APIs by writing shims manually, which is tedious and can be error-prone, and, as a result, can risk soundness.
More details about how we model them are explained in 
\techref{Appendix~\ref{app:side}.}{our technical report~\cite{supp}.}

\myparagraph{Communication Message Format} Based on the above analysis, we can record the ``process communication'' as observable effects for each code fragment. The detailed grammar for expressing these communication messages is listed in Fig. \ref{fig:grammarsem}.
The whole recording (\textbf{Semantics} in Fig. \ref{fig:grammarsem}) contains sets of message sequences ($\overrightarrow{\textbf{MsgSeq}}$) involving each code fragment (identified by $\textbf{Id}^\text{Scope}$).
Each \textbf{MsgSeq} represents the message sequence ($\overrightarrow{\textbf{IOStep}}$) in and out of one process, corresponding to a single execution of a code fragment.
There are 3 types of \textbf{Input} messages and 3 types of \textbf{Output} messages for a fragment process.
Each message has associated data, including a concise context \textbf{CTX} determined by the aforementioned analysis of may-access objects as well as control-flow specific data transfer (such as \textbf{ARGS} in the {\sf Call} message).

Fig. \ref{fig:ovsemantics} demonstrates the observable effects captured in our \procint model for a single execution of the code fragment \code{compute} in our example program.
This execution of \code{compute} corresponds to process $P_1$ in the earlier Fig. \ref{fig:workflowstepmodel}.
The middle of Fig. \ref{fig:ovsemantics} shows how $P_1$ communicates with the skeleton process $\mathcal{K}_\src$ in our conceptual \procint model design.
Our analysis to obtain observable effects related to process $P_1$ will be equivalent to logging the messages when executing the pseudo-code shown in the middle. 
With similar analysis for other fragments, our analyzer can collect the observable effects as shown on the right of Fig. \ref{fig:ovsemantics}.

\input{figtabs/fig_ov_semantics_type}

\subsubsection{Transferring the Observable effects to the Target}\label{sec:semstep2}  

After obtaining the observable effects from the source \py program, the next step is to convert those observable effects into semantic requirements for the placeholders in the target skeleton.

\todoadd{The main objective of the conversion is to map data types.
Most parts of the representation in \decocr (corresponding to \textbf{Semantics} in Fig. \ref{fig:grammarsem}) are kept as is during conversion.}
The only changing places are the data type names (i.e., $\textbf{Id}^\text{type}$) in the data object representation (i.e., \textbf{VAL}).
 For example, we change all the $\textbf{Id}^\text{type} = ``\texttt{List}"$ (a type in \py) into $\textbf{Id}^\text{type} = ``\texttt{Array}"$ (a type in \js).
The semantic contents and relations of data objects are kept untouched.

\myparagraph{Type Mapping} \todoadd{Choosing which data type to map to in the target language is an important problem that \name has to address.}
Idiomatic translations may often use semantically similar types, but for every source data type, there is no single target data type that is universally the best choice in all translation tasks.
As an example, Fig. \ref{fig:typemapping} shows details of the same \py code fragment (\code{compute}) and an example \js translation.
The variable \code{arr} (highlighted in Fig. \ref{fig:typemapping}) in the \py code fragment refers to objects of type \code{List[int]} during execution, which can be seen from its observable effects highlighted \ignore{in green}in Fig. \ref{fig:typemapping}.
In the translated code fragment (on the right of Fig. \ref{fig:typemapping}), the \code{arr} refers to objects of type \code{Array<number>}, which is a commonly used type in \js.
\ignore{
The type \code{number} in \js is not exactly equivalent to \code{Int} in \py since their value ranges are different.
Nevertheless, the actual semantic content of \code{arr} remains the same across \py and \js.
}
However, many other choices exist as well.
Alternative types such as \code{Array<BigInt>} and \code{BigInt64Array} in \js can support larger integer values 
or memory-efficient operations.

\name uses a default type mapping $\mathcal{F}$ (partially shown in the middle of Fig. \ref{fig:typemapping}) that is context insensitive when transferring the observable effects into semantic requirements in the skeleton. 
Type consistency is guaranteed by $\mathcal{F}$ since we can determine, for example, that for every \py object with type \code{List[int]}, the corresponding object in \js must have type \code{Array<number>}.
The default mapping can be overwritten if needed to be tailored to specific translation tasks. Automatically deciding the best type mapping (potentially context-sensitive) for a translation task is orthogonal to \name and can be useful future work.

\myparagraph{Translation Flexibility}
The ability to abstract away details with observable effects allows \name to have flexible translations. 
For example, the \code{zip(mat, arr)} function call in the \py code fragment in Fig.~\ref{fig:typemapping}  creates a stateful iterator object, which is updated multiple times before calls to function \code{multi}.
However, this iterator object used in \code{compute} is never accessed by the rest of the program, and is thus omitted from the observable effects.
The corresponding \js translation is free to choose how the loop is implemented, as long as observable effects are the same (such as the same sequence of {\sf Call} messages, etc.). 
For example,  Fig. \ref{fig:typemapping} shows a \js translation on the right. The object corresponding to the \code{zip} iterator in \py is gone, and the loop is expressed using the \code{forEach} API from the standard library in \js, which is internally stateful.

%% file: figtabs/figworkflow-stepmodel.tex
\begin{figure}[ht]
    \centering
    \includegraphics[trim=1.3in 1.8in 1.3in 1.8in, clip, width=0.95\linewidth]{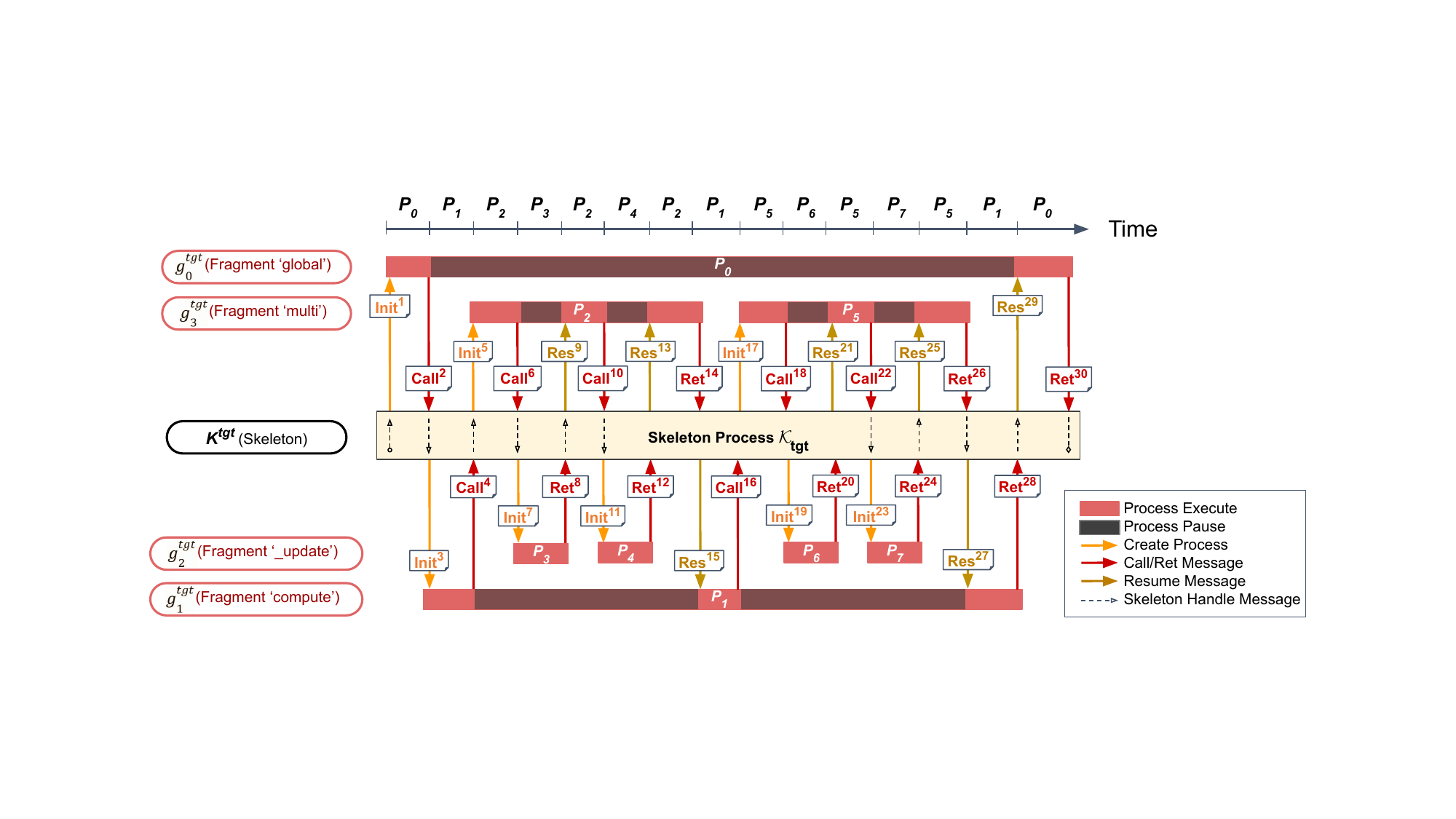}
    \caption{A conceptual multi-process execution trace when executing the translation (or the corresponding source) in our \procint design. Each invocation of a code fragment runs in a separate ``process''.}
    \label{fig:workflowstepmodel}
\end{figure}

%% file: figtabs/defsem.tex
\begin{figure}[!htbp]
    \centering
    \footnotesize
    \begin{tabular}{@{}r@{\hskip6pt}c@{\hskip6pt}p{5.5cm}p{6cm}@{}}
        \textbf{\decocr} & ::= & $(\textbf{Syntax}, \textbf{Semantics})$ & // The grammar of {\decocr} \\
        \textbf{Semantics} & ::= & $\textbf{Id}^{\text{Scope}} \to \overrightarrow{\textbf{MsgSeq}}$ & // Each scope has a set of message sequences \\

        \textbf{MsgSeq} & ::= & $\overrightarrow{\textbf{IOstep}}$ & // Each MsgSeq is a sequence of Input/Output steps \\

        \textbf{IOstep} & ::= & $(\textbf{Input}, \textbf{Output})$ & // Each I/O step contains Input and Output \\

        \textbf{Input} & ::= & Init(\textbf{CTX})& // Initialize execution \\
        
        & & | Resume(\textbf{CTX}, $\textbf{Id}^\text{Obj}$) | ResumeThr(\textbf{CTX}, $\textbf{Id}^\text{Obj}$) & // Resume execution from Return/Throw \\

        \textbf{Output} & ::= & Call($\textbf{CTX}, \textbf{VAL}, \overrightarrow{\textbf{Id}^\text{Obj}}$) & // Closure/function call (\textbf{VAL} is a closure) \\
        & & | Return($\textbf{CTX}, \textbf{Id}^\text{Obj}$) | Throw($\textbf{CTX}, \textbf{Id}^\text{Obj}$)  & // Return/Throw back to the caller process \\
        
        \textbf{CTX} & ::= & $(\textbf{VARS}, \textbf{OBJECTS})$ & // Execution context \\
        
        \textbf{VARS} & ::= & $[\textbf{Id}^\text{Var} \to \textbf{Id}^\text{Obj}]$ & \text{// Obserable variables} \\
        
        \textbf{OBJECTS} & ::= & $[\textbf{Id}^\text{Obj} \to \textbf{VAL}]$ & \text{// Obserable objects} \\

        \textbf{VAL} & ::= & \text{Collection}($\textbf{Id}^\text{type}$, $\overrightarrow{\textbf{Id}^{\text{Obj}}}$) | \text{Primitive}($\textbf{Id}^\text{type}$, \text{Val})& \text{// Data Values (Collection/Primitive types)} \\
        & & | \text{Closure}($\textbf{Id}^\text{Scope}, \textbf{Id}^\text{Proc}$) & // Data Values (Closure type)\\
   
    \end{tabular}

    \caption{The remaining part of \decocr's grammar that represents \procint semantics (continued from Fig.~\ref{fig:grammarsyn}).}
    \label{fig:grammarsem}
\end{figure}

%% file: figtabs/fig_ov_semantics.tex
\begin{figure}[t]
    \centering
    \includegraphics[trim=1.9in 1.2in 1.9in 2.1in, clip, width=0.85\linewidth]{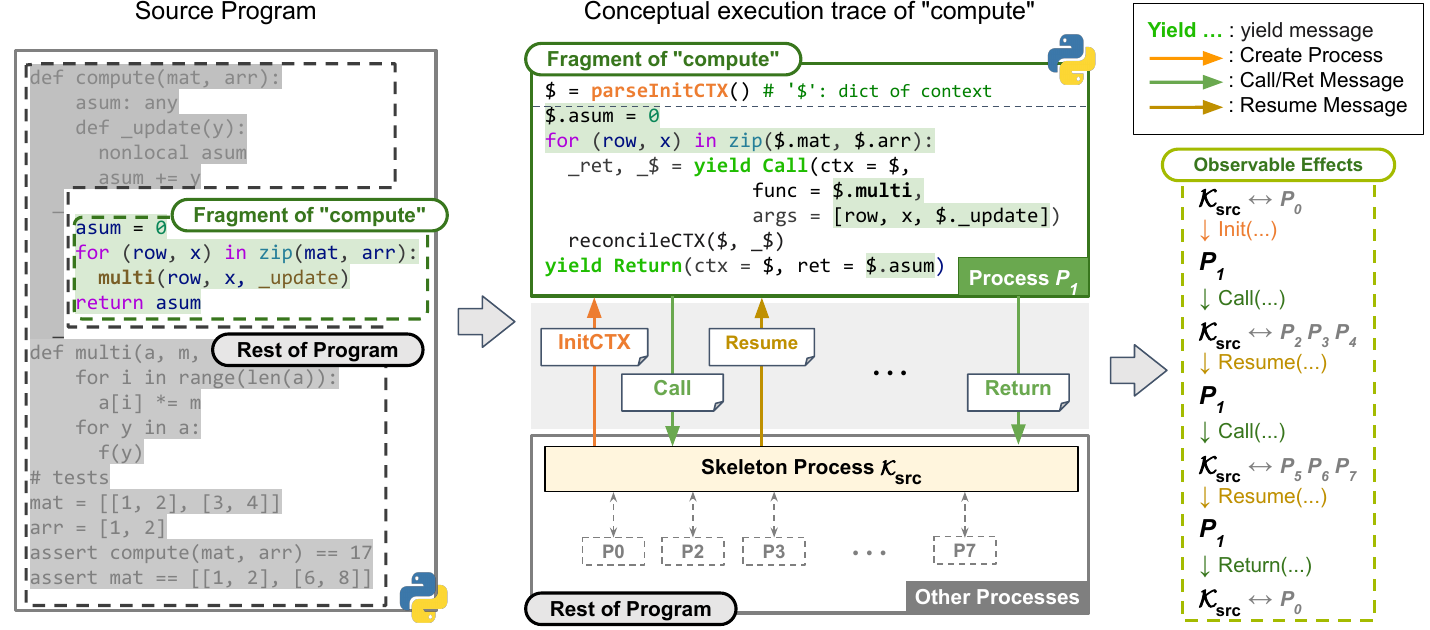}
    \caption{A conceptual view of the observable effects of the code fragment \code{compute} under our \procint model. The middle demonstrates the communication messages. The actual semantics of the original program (left) maps to the \procint semantics (right) as observable effects for the code fragment.}
    \label{fig:ovsemantics}
\end{figure}

%% file: figtabs/fig_ov_semantics_type.tex
\begin{figure}[t]
    \centering
    \includegraphics[trim=1.4in 0.9in 1.4in 2.25in, clip, width=0.97\linewidth]{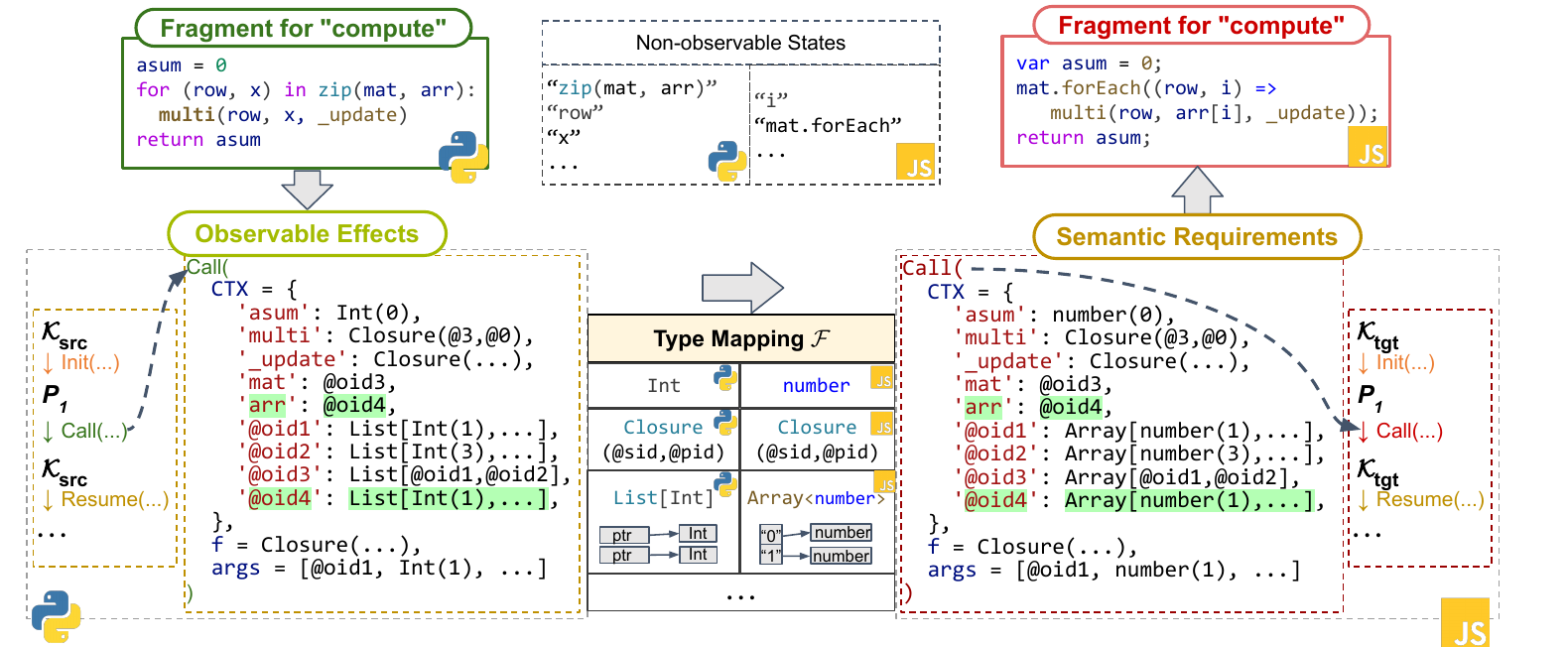}
    \caption{Type Mapping used by \name maps observable effects of code fragments in the Python source into semantic requirements of placeholders in the target program skeleton.}
    \label{fig:typemapping}
\end{figure}

%% file: sections/4_3_synth.tex
\subsection{Code Fragment Synthesis}\label{sec:codesynth}

After obtaining the program skeleton, \name synthesizes code fragments for placeholders.
Examples of such synthesized fragments are highlighted as 4 dashed boxes in the \js translation of Fig.~\ref{fig:ovsyntax}, corresponding to $g_0^\tgt,..., g_3^\tgt$. These fragments combined with the program skeleton in the target language $K^\tgt$ become the complete translation. This step in \name uses external synthesizers.

\name synthesizes and refines those code fragments following the program execution order step-by-step. Each fragment is generated by a code synthesizer (e.g., LLMs).
We propose an algorithm termed the Execution-Order Translation (\jitt) loop to handle the whole process.
The \jitt loop uses \procint to check the incomplete translation after every step.
If the current code fragment to be executed is missing, the \jitt loop queries a synthesizer to obtain an initial code fragment. 
Otherwise, if a code fragment already exists, the \jitt loop applies \textit{check-and-refine} strategy on this fragment. 
Execution steps that already pass the check are skipped, and only counterexamples (i.e., the steps that fail the check) will be provided to the synthesizer to refine that code fragment.
Once the \jitt loop terminates, the final translation naturally passes all tests from which the semantic requirements are derived.
Next, we explain the \jitt loop with an example in detail. The precise algorithm for \jitt is presented in 
\techref{Appendix~\ref{app:eot}.}{our technical report~\cite{supp}.}

\myparagraph{\jitt Illustration}
Fig. \ref{fig:workflowstepjit} demonstrates an example run of the \jitt loop that aims to fill the target skeleton such that the completion realizes the execution sequence $\rho$ shown earlier in Fig. \ref{fig:workflowstepmodel}.
The intermediate steps involving the skeleton process $\mathcal{K}$ are omitted for simplicity.
There are three possible cases when processing each step: ``Missing Fragment'', ``Step Error'', and ``Step Pass''.

\input{figtabs/figworkflow-stepjit}

``Missing Fragment'' means that the fragment behind the conceptual process $P_i$ at this step has not yet been implemented, e.g., the second step (involving process $P_1$) in the figure. 
\todoadd{The \jitt loop will construct a query for synthesizing this code fragment.
We use the word ``specification'' to refer to input-output steps provided to the synthesizer, which will be a chosen subset of the semantic requirements for each code fragment, independent from other code fragments.
When ``missing fragment'' happens, the \jitt loop will select the first input-output step involving that fragment as the initial specification.
The initial specification, together with the corresponding source code fragment (in \py) as a hint, are combined as one query to the synthesizer (i.e., LLM).}

The second and third possible cases happen on ``processes'' that already have corresponding code fragments translated. 
If the existing code fragment behaves as expected for the step, it is a ``Step Pass'', and the \jitt loop will move on to the next step in $\rho$. 
Otherwise, it is a ``Step Error''. The \jitt loop will either refine the specification (i.e., adding counterexamples) or repair the code fragment (i.e., retry with the error message provided). As an example, the second-to-last step in Fig. \ref{fig:workflowstepjit} shows an error step when executing $P_1$, triggering a refinement for the code fragment \code{compute}.

The refine and repair procedure aims to iteratively update the code fragment to pass the error steps, while not breaking any earlier steps it has passed. 
It is conceptually irrelevant to other code fragments in the program because the root cause for the error step is guaranteed to be within the code fragment to be updated.
To fix the error, the \jitt loop will first refine the specification by adding the error step as a counterexample. 
Then, a new code fragment will be synthesized, and all execution steps involving this code fragment will be executed and validated again.
If all the validation checks pass, this error is considered resolved, and the \jitt loop will move forward.
For example, the error step in Fig. \ref{fig:workflowstepjit} is resolved by refining the specification to add one counterexample for the code fragment \code{compute}.

It is also possible that when the code fragment is updated, some previous steps involving the same code fragment report an error. 
If this previous step is not yet selected as a counterexample, the specification for the code fragment will be refined to include it. %
Another possible situation is that the code fragment's behavior directly violates the specification accumulated so far for that code fragment. In this case, \name will 
highlight the mismatch and
instruct the synthesizer to repair the code it has generated. More specifically, \name provides to the LLM a description of the semantic mismatch together with the code fragment to fix the error in the code.
This iterates until either the error step is resolved or reaches a pre-determined retry limit for repair.
In the latter case, the \jitt loop will pause and wait for external assistance, since the synthesizer might not be capable of solving it.
Finally, after processing all steps in $\rho$, we will obtain a full translation that is correct on tests, such as the one shown in Fig. \ref{fig:ovsyntax}.

%% file: figtabs/figworkflow-stepjit.tex
\begin{figure}[ht]
    \centering
    \includegraphics[trim=1.4in 1.6in 1.5in 2.3in, clip, width=0.9\linewidth]{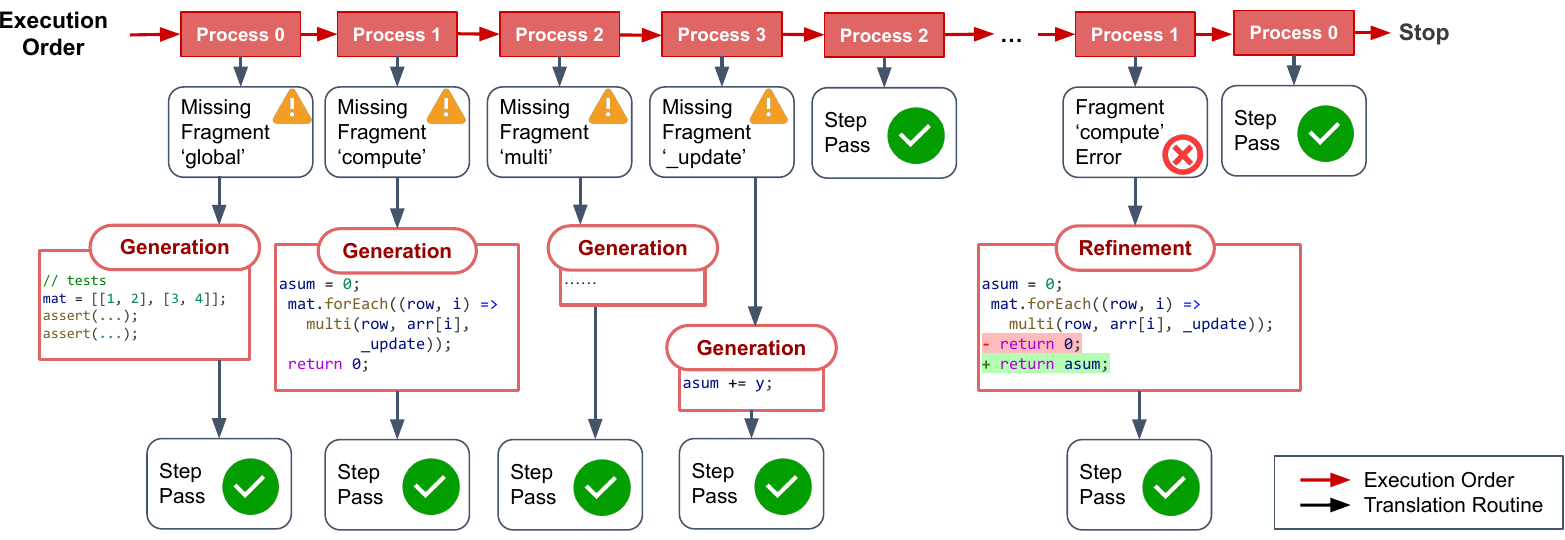}
    \caption{The Execution-order Translation (\jitt) loop synthesizes, checks, and refines fragments.
    }
    \label{fig:workflowstepjit}
\end{figure}

%% file: sections/6impl.tex
\section{Implementation}
\label{sec:impl}

Here we explain the implementation of \name prototype that follows our design in Section \ref{sec:design}.

\myparagraph{Supported Language Subset}
Our \name prototype focuses on subsets of \py and \js whose programs satisfy the following properties, making them amenable to being modeled in our specific \procint model.
First, programs in these subsets are deterministic and single-threaded.
Second, for these programs, lexical scopes cannot be created or modified during runtime, and cannot be accessed without using variables in the symbol table.
This rules out programs using reflection (e.g., \code{eval(..)}).
Third, we assume that most of the library APIs and operators used in these programs are either pure
or result in transparent effects, as explained earlier in Section \ref{sec:skeletonsemantics}.
More than 90\% of the APIs that we observe in our experimental benchmarks satisfy these criteria. %
For a handful of APIs that have opaque side effects (e.g., \code{random} and \code{print}),
we implemented an API shim to compute their effects as process communication messages under \procint. The implementation of such API shim is currently best-effort and might be incomplete. More details are provided in \techref{Appendix~\ref{app:side}.}{our technical report~\cite{supp}.}

\myparagraph{Rewriting \pybold Language Features}
We only model closures and positional arguments in the design of \decocr.
To support programs consisting of classes, decorators, keyword arguments, etc., our \name prototype rewrites them to nested closures and positional arguments by a simple normalization step.
Class inheritance and class methods are also handled by such a normalization step.
Details of our normalization strategies are provided in 
\techref{Appendix~\ref{app:norm}.}{our technical report~\cite{supp}.}

\myparagraph{Skeleton Generation}
The syntactic structure (expressed in \decocr) is extracted through lightweight analysis on the source to resolve symbol definitions and references.
The analysis eliminates local symbols that are not referred to elsewhere and symbol tables for \todoadd{small} nested closures that for sure will not escape.
For semantic requirements of the skeleton, we instrument all possible control-flow transfers in and out of each code fragment.
The dynamic analyzer will also keep track of metadata for shared objects and closures \todoadd{as explained in Section \ref{sec:semstep1}.} \tododel{To analyze shared objects as mentioned in Section \ref{sec:semstep1}, 
our analyzer determines whether data objects may be externally observable or internal, based on the following rule: 
any objects passed through the messages between ``processes'' will be marked as observable for both processes and remain observable for them until they finish.}

%% file: sections/7eval.tex
\section{Evaluation}
\label{sec:eval}

We evaluate \name for the task of translating programs from \py to \js on two aspects:

\begin{enumerate}
    \item \textbf{Effectiveness} (Section 5.1): Does \name translate real-world programs from \py to \js mostly automatically with reasonable correctness? 
    \item \textbf{Ablation Study} (Section 5.2): How much does each component of \name contribute? 
\end{enumerate}

\myparagraph{Benchmarks}
We expect each sample program in our benchmark to contain a standalone \py program with tests. 
One existing benchmark satisfying such requirements is available from previous work on debugging neural translations~\cite{transmap}. This benchmark consists of $5$ real-world \py modules with sizes varying from $\progsizeJJJtransmapmin$ to $\progsizeJJJtransmapmax$ LOC (excluding tests).
In addition to this benchmark, we collect 4 longer and more diverse programs from popular GitHub repositories with at least 700 stars. 
These comprise our benchmarks shown in Table~\ref{tab:benchmark}. There are 9 programs in total ranging from $121$ LOC to $2400$ LOC. 
They range from implementations of classical algorithms to modules in the \py standard library or third-party modules used in production. 
We report the maximum depth of its static call graph for each program ($h_\text{CG}$ column). We see that longer programs tend to have larger $h_\text{CG}$, indicating more complexity in program structures. 
\ignore{It is worth pointing out that \texttt{py-evtx} has a much higher $h_\text{CG}=26$ compared with other programs with $h_\text{CG}$ between $2$ and $8$.}

\input{figtabs/tabs/tabbenchmark}

To reduce the difficulty for the external LLM synthesizer, we set a limit for the maximum size of a single code fragment to be 100 lines.
All but 5 functions in our benchmark programs are already shorter than 100 LOC.
We syntactically refactor the body of those 5 functions into smaller functions shorter than 100 LOC.
This step is fairly straightforward and is automatable with existing IDE tools.
We also combine code files (if there are multiple) into a self-contained single-file \py program.
The statistics shown in Table \ref{tab:benchmark} are computed on such preprocessed programs, which will be used for evaluating \name as well as comparing \name with baseline approaches.

\myparagraph{Unit Tests\todoadd{~and Global Data}} 
Each of the benchmark \py programs comes with existing test suites. \todoadd{We extend some of the test suites to increase the code coverage, and we use our extended test suites for evaluation. The final test coverage is listed in Table \ref{tab:benchmark}.}
We use simple string replacement (following previous work~\cite{unsupervised,transmap}) to convert \py tests into \js tests. 
This suffices since the tests are often routine sequences of calls and value comparisons, which do not require idiomatic code synthesis. 
We manually check them to ensure correctness, and \name also helps confirm that they behave the same across the source and target. \todoadd{We also did similar conversions for global constants in the program, which include array initializers or lookup tables that are potentially large in size but straightforward to convert.}
\ignore{that since wrong tests can cause \name to stop, reporting that the test code fragment has different observable effects from the \py source.}

\myparagraph{\name Setup}
We configure \name to use the same default type mapping for the translation of all programs. 
For the \jitt loop in \name, we set the maximum retry limit for repair as 3. 
It means that the \jitt loop will pause and wait for external assistance if an execution step 
is still failing after 3 rounds of retrying.
We use ``GPT-4-turbo'' and ``GPT-3.5-turbo''\footnote{We use the gpt-4-turbo-2024-04-09 and gpt-3.5-turbo-0125 versions.}, which are among the state-of-the-art LLM models, for the LLM-based code synthesis and prompt-based repair. 
We set the decoding temperature to 0 to reduce noise in the output.
We fix a prompt template for the evaluation of all programs. The details of the prompts are listed in 
\techref{Appendix~\ref{app:prompt}.}{our technical report~\cite{supp}.}

\myparagraph{Compared Translators} We compare \name with two baselines, one based on LLMs and one based on compiler rules. 
The first baseline (LLM-based) is a simple syntactic divide-and-conquer strategy we implemented using the same model, same hyperparameters, and similar prompts as \name. 
We sequentially divide the source into segments where each is a complete function or class definition in the global scope.
We query the LLM to translate each segment and concatenate the translations back.
The second baseline (rule-based) is an existing compiler-based translator used in production, called Transcrypt \cite{transcrypt}.
It is one of the most developed rule-based translators for \py to \js translation, with around $3k$ stars on GitHub.

\myparagraph{Evaluation Metrics} 
We use two correctness metrics with different granularity.
The first one is the \emph{correctness} of the whole translation, determined by whether it passes the tests.
\ignore{The first one is whether the whole program passes the tests.
If it does, we say that the translation is \emph{correct}.}
To better compare the result when we do not automatically get a correct translation, we use the second metric: the number of functions that need external assistance (e.g., from a user) to fix. We name it \textbf{\#UserFix} for short.
For \name, \textbf{\#UserFix} is the number of functions that \name's \jitt loop will stuck on (exhausted the retry limit). Human intervention can provide a correct code fragment to fill the placeholder so that \name can continue with the rest of the translation.
For baselines, \textbf{\#UserFix} is not easy to determine since there is no automated step-by-step validation like \name.
Thus, we make a best-effort attempt to start debugging and fixing the most obvious errors in the translation, such as type errors, non-existing APIs, and so on. If a fix is inside a function, we count that function in \textbf{\#UserFix}. We stop after spending 1-2 hours for each program to obtain a \textbf{\#UserFix} lower-bound reported as ($k+$), if the program is still not correct after fixing $k$ functions.

\subsection{Overall Effectiveness}
\label{sec:evaleffective}

We first evaluate the overall effectiveness of \name equipped with either GPT-4-turbo or GPT-3.5-turbo as the code synthesizer. 
Since \name guarantees to pass the tests as long as \jitt loop terminates successfully,
our evaluation is to run \name to translate those 9 programs and provide human interventions when \name is stuck at a code fragment. 
We count the number of functions that need intervention (\textbf{\#UserFix}).
Eventually, the translations pass the tests, as expected by our design. We report the number of functions that are translated and validated by \name without human intervention as \textbf{\#Auto}.
We show the numbers (i.e., \textbf{\#Auto} and \textbf{\#UserFix}) in Table \ref{tab:deco}. We also use an automation ratio to represent the ratio of \textbf{\#Auto} / (\textbf{\#Auto} + \textbf{\#UserFix}).

We find that \name with GPT-4 can automatically translate \textbf{\skelfourAutoProgCount} out of 9 real-world programs (highlighted in \highlightgreen{green}) without human intervention. 
In more detail, \name with GPT-4 can automatically translate \autogptfourJJJTOTAL~ of \countfJJJTOTAL~ functions\footnote{\todoadd{We count the number of functions in the normalized source program that are covered by tests (but exclude tests themselves). We omit wrapper functions created when normalizing classes since they are mechanically generated as part of the skeleton.}} and reach an overall\tododel{ \percentskelauto\% of} automation ratio\todoadd{~of around \percentskelauto\%}. 
With a weaker code synthesizer (GPT-3.5), \name can only automatically translate \skelthreeAutoProgCount~program, but the overall automation ratio still reaches \todoadd{around} \autogptthreepctJJJTOTAL\%.
All the final translations pass the tests.

\begin{tcolorbox}[breakable,width=\linewidth,boxrule=1pt,top=0pt, bottom=0pt, left=1pt,right=1pt]
The effectiveness of \name improves when a stronger synthesizer is used. \name equipped with GPT-4 automatically translates $\percentskelauto$\% functions correctly. Final translations all pass the tests.
\end{tcolorbox}

Then we compare the effectiveness of \name and two baseline translators. Here, we use GPT-4 as the LLM baseline. Table~\ref{tab:baseline} shows the number of \textbf{\#UserFix} functions in the translation of each benchmark. $0$ means that the translation is correct without any human intervention. As explained in \textit{Evaluation Metrics}, we report the lower bound of \textbf{\#UserFix}  (annotated in $k+$) if we (authors) cannot fix the translation after a limited amount of effort and the translation still fails on tests.

\input{figtabs/tabs/tab2skelperformance.tex}

\input{figtabs/tabs/tab3compare.tex}

As shown in Table~\ref{tab:baseline}, none of the $9$ benchmarks can be correctly translated by the baseline LLM approach.  
From the perspective of \textbf{\#UserFix} functions, more than $\fixgptfourbaseJJJTOTALTRIM$ functions in the translations by the LLM baseline approach need to be fixed. This process is time-consuming and tedious, especially for the two programs longer than $1k$ LoC. We (authors) have spent more than dozens of hours fixing the translations but still failed to make $\gptfourbaseFailProgCount$ programs pass all tests. 
The rule-based translator Transcrypt can correctly translate $\transcryptbasePassProgCount$ programs.
However, it does not run on $\transcryptbaseNAProgCount$ other programs (annotated with NA) because these programs use APIs and language features unsupported by Transcrypt.
The process of fixing the translations by Transcrypt is also not easy. Its translations use emulated libraries and dependencies by Transcrypt and have poor readability.
For comparison, to reach fully correct translations on 9 programs, \name with GPT-4 requires around \textbf{\userfixRatioSkelVSBase} of \textbf{\#UserFix} compared with the baseline approach. 
In the meantime, with the help of the step-by-step checking in \name, the location of the error can be located down to the scope of a single code fragment, making it much easier for human intervention.
We also highlight that \name equipped with a weaker GPT-3.5 model can still perform better than the baseline approach equipped with GPT-4.

\begin{tcolorbox}[breakable,width=\linewidth,boxrule=1pt,top=0pt, bottom=0pt, left=1pt,right=1pt]
The translation by \name has much fewer \textbf{\#UserFix} functions compared with other translators. Step-by-step checks in \name also make it easy to tell where to fix when the user intervenes.
\end{tcolorbox}

\subsection{Ablation Study}
\label{sec:evalablation}

We conduct an ablation study to evaluate how much each design choice in \name's code synthesis mechanism contributes to its level of automation, which in turn reduces human effort.
For example, the most basic approach to filling the program skeleton may only provide the corresponding source code fragment in syntax, 
without providing the semantic requirements that are used to validate each code fragment.
We empirically validate the necessity of two choices in the code synthesis process to the final automation ratio: (a) providing one step of semantic requirements as \emph{specifications} besides the source code fragment; and (b) \emph{check-and-refine}, which aims to automatically validate every step and refine and fix the code when the synthesizer (LLM) cannot get it correct in the first try.
We use $\text{\name}_\text{base}$, $\text{\name}_\text{spec}$, and $\text{\name}_\text{spec+chkrfn}$ to represent the \name working without these two (only the source code fragment is provided), \name with the semantic specification in synthesis, and the complete \name with both semantic specification and stepwise check-and-refine, respectively. 
We test each variant of \name with GPT-4 serving as the code synthesizer and report the number of \textbf{\#UserFix} for each \name variant. 
The results are shown in Table \ref{tab:ablation}.

\input{figtabs/tabs/tab4ablation.tex}

Without the help from semantic specifications and iterative refinement, $\text{\name}_\text{base}$ produces \skelbaseJJJTOTAL ~mistaken functions during the translation. After adding the semantic specification into \name, about 1/3 of \textbf{\#UserFix} is no longer needed, and $\skelspecJJJTOTAL$ \textbf{\#UserFix} remain. 
This shows that the semantic specifications help clarify the task for LLMs, but the synthesized code is still error-prone.
After adding the iterative refinement into \name, another \ablationGapSpecMinusChkrfn ~\textbf{\#UserFix} can be automated.
With the help of check-and-refine, LLMs can eventually synthesize a correct code fragment in most cases. 
A further discussion on the remaining \textbf{\#UserFix} is in 
\techref{Appendix~\ref{app:fail}.}{our technical report~\cite{supp}.}

%% file: figtabs/tabs/tabbenchmark.tex
\newcommand{\countfunc}{\#\text{Function}}
\newcommand{\hcg}{h_\text{CG}}
\newcommand{\locf}{\overline{\text{LoC}_\text{F}}}
\newcommand{\loccover}{\text{Coverage}}
\begin{table}[h]
\footnotesize
    \centering
    \setlength{\abovecaptionskip}{1ex}
    \caption{
    Summary of our benchmarks. The LOC / SLOC column shows the Lines of Code with and w/o comments. $\hcg$ is the maximum depth of the call graph. $\loccover$ denotes the line coverage achieved by the unit tests. $\locf$ represents the average lines of code per function.}
    \label{tab:benchmark}
    \begin{tabular}{l|ccccl}
        \toprule
         Program name & LOC / SLOC & $\hcg$ & $\loccover$ & $\locf$ & Description \\
         \midrule
         \texttt{colorsys} & \progsizeLocJJJcolorsys / \progsizeSlocJJJcolorsys & \hcgJJJcolorsys & \covJJJcolorsys\% & \locfJJJcolorsys & Color conversion\\
         
         \texttt{heapq} & \progsizeLocJJJheapq / \progsizeSlocJJJheapq & \hcgJJJheapq & \covJJJheapq\% & \locfJJJheapq & Heap data structure\\

         \texttt{html} & \progsizeLocJJJhtml / \progsizeSlocJJJhtml & \hcgJJJhtml & \covJJJhtml\% & \locfJJJhtml & HTML parser\\
         
         \texttt{mathgenerator} & \progsizeLocJJJmathgen / \progsizeSlocJJJmathgen & \hcgJJJmathgen & \covJJJmathgen\% & \locfJJJmathgen & Math question generator\\

         \texttt{strsimpy} & \progsizeLocJJJstrsim / \progsizeSlocJJJstrsim & \hcgJJJstrsim & \covJJJstrsim\% & \locfJJJstrsim & String distance and similarity\\

         \texttt{bst-rec} & \progsizeLocJJJbsdrec / \progsizeSlocJJJbsdrec & \hcgJJJbsdrec & \covJJJbsdrec\% & \locfJJJbsdrec & Binary search tree\\
        
         \texttt{red-black-tree} & \progsizeLocJJJrbtree / \progsizeSlocJJJrbtree & \hcgJJJrbtree & \covJJJrbtree\% & \locfJJJrbtree & Red-black tree\\
                  
         \texttt{toml} & \progsizeLocJJJtoml / \progsizeSlocJJJtoml & \hcgJJJtoml & \covJJJtoml\% & \locfJJJtoml & Parser for TOML \\

         \texttt{py-evtx} & \progsizeLocJJJpyevtx / \progsizeSlocJJJpyevtx & \hcgJJJpyevtx & \covJJJpyevtx\% & \locfJJJpyevtx & Parser for Windows event logs\\
        \bottomrule
    \end{tabular}
\end{table}

%% file: figtabs/tabs/tab2skelperformance.tex
\begin{table}[t!]
\footnotesize
    \centering
    \setlength{\abovecaptionskip}{1ex}
    \caption{The level of automation of translations by \name with different code synthesizers. The table shows the number and percentage of \#Auto-translated functions and \#UserFix functions. 
    Programs that are translated without and \#UserFix is marked in in green.
    }
    \label{tab:deco}
    \begin{tabular}{l|c|c|c|c|c}
        \toprule
            \multirow{2}{*}{\textbf{Programs}} & \multirow{2}{*}{\#Function} & \multicolumn{2}{c|}{\name with GPT-4} & \multicolumn{2}{c}{\name with GPT-3.5}\\
        \cmidrule{3-4}
        \cmidrule{5-6}
        &  & \#Auto (\%) & \#UserFix (\%) & \#Auto (\%) & \#UserFix (\%)\\
            \midrule
            colorsys & \countfJJJcolorsys & \cellcolor{lime!30}\autogptfourJJJcolorsys\ (\autogptfourpctJJJcolorsys\%) & \fixgptfourJJJcolorsys\ (\fixgptfourpctJJJcolorsys\%) & \autogptthreeJJJcolorsys\ (\autogptthreepctJJJcolorsys\%) & \fixgptthreeJJJcolorsys\ (\fixgptthreepctJJJcolorsys\%)\\
            
            heapq & \countfJJJheapq & \autogptfourJJJheapq\ (\autogptfourpctJJJheapq\%) & \fixgptfourJJJheapq\ (\fixgptfourpctJJJheapq\%) & \autogptthreeJJJheapq\ (\autogptthreepctJJJheapq\%) & \fixgptthreeJJJheapq\ (\fixgptthreepctJJJheapq\%)\\

            html & \countfJJJhtml & \autogptfourJJJhtml\ (\autogptfourpctJJJhtml\%) & \fixgptfourJJJhtml\ (\fixgptfourpctJJJhtml\%) & \autogptthreeJJJhtml\ (\autogptthreepctJJJhtml\%) & \fixgptthreeJJJhtml\ (\fixgptthreepctJJJhtml\%)\\
            
            mathgen & \countfJJJmathgen & \autogptfourJJJmathgen\ (\autogptfourpctJJJmathgen\%) & \fixgptfourJJJmathgen\ (\fixgptfourpctJJJmathgen\%) & \autogptthreeJJJmathgen\ (\autogptthreepctJJJmathgen\%) & \fixgptthreeJJJmathgen\ (\fixgptthreepctJJJmathgen\%)\\

            strsim & \countfJJJstrsim & \cellcolor{lime!30}\autogptfourJJJstrsim\ (\autogptfourpctJJJstrsim\%) & \fixgptfourJJJstrsim\ (\fixgptfourpctJJJstrsim\%) & \cellcolor{lime!30}\autogptthreeJJJstrsim\ (\autogptthreepctJJJstrsim\%) & \fixgptthreeJJJstrsim\ (\fixgptthreepctJJJstrsim\%)\\

            bst-rec & \countfJJJbsdrec & \cellcolor{lime!30}\autogptfourJJJbsdrec\ (\autogptfourpctJJJbsdrec\%) & \fixgptfourJJJbsdrec\ (\fixgptfourpctJJJbsdrec\%) & \autogptthreeJJJbsdrec\ (\autogptthreepctJJJbsdrec\%) & \fixgptthreeJJJbsdrec\ (\fixgptthreepctJJJbsdrec\%)\\

            red-black-tree & \countfJJJrbtree & \cellcolor{lime!30}\autogptfourJJJrbtree\ (\autogptfourpctJJJrbtree\%) & \fixgptfourJJJrbtree\ (\fixgptfourpctJJJrbtree\%) & \autogptthreeJJJrbtree\ (\autogptthreepctJJJrbtree\%) & \fixgptthreeJJJrbtree\ (\fixgptthreepctJJJrbtree\%)\\
            
            toml & \countfJJJtoml & \autogptfourJJJtoml\ (\autogptfourpctJJJtoml\%) & \fixgptfourJJJtoml\ (\fixgptfourpctJJJtoml\%) & \autogptthreeJJJtoml\ (\autogptthreepctJJJtoml\%) & \fixgptthreeJJJtoml\ (\fixgptthreepctJJJtoml\%)\\

            py-evtx & \countfJJJpyevtx & \autogptfourJJJpyevtx\ (\autogptfourpctJJJpyevtx\%) & \fixgptfourJJJpyevtx\ (\fixgptfourpctJJJpyevtx\%) & \autogptthreeJJJpyevtx\ (\autogptthreepctJJJpyevtx\%) & \fixgptthreeJJJpyevtx\ (\fixgptthreepctJJJpyevtx\%)\\

        \midrule
                    
            total & \countfJJJTOTAL & \autogptfourJJJTOTAL\ (\autogptfourpctJJJTOTAL\%) & \fixgptfourJJJTOTAL\ (\fixgptfourpctJJJTOTAL\%) & \autogptthreeJJJTOTAL\ (\autogptthreepctJJJTOTAL\%) & \fixgptthreeJJJTOTAL\ (\fixgptthreepctJJJTOTAL\%)\\
        \bottomrule
    \end{tabular}
\end{table}

%% file: figtabs/tabs/tab3compare.tex
\begin{table}[th!]
\footnotesize
    \centering
    \setlength{\abovecaptionskip}{1ex}
    \caption{Numbers of \textbf{\#UserFix} functions compared with 2 baselines.}
    \label{tab:baseline}
    \begin{tabular}{l|c|c|c|c|c}
        \toprule
            \textbf{Programs} & \#Function & Baseline with GPT-4 & Transcrypt & \name with GPT-4 & \name with GPT-3.5 \\
            
            \midrule
            colorsys & \countfJJJcolorsys & \fixgptfourbaseJJJcolorsys & \cellcolor{lime!30}\fixtranscryptJJJcolorsys & \cellcolor{lime!30}\fixskelgptfourJJJcolorsys & \fixskelgptthreeJJJcolorsys\\
            
            heapq & \countfJJJheapq & \fixgptfourbaseJJJheapq & \fixtranscryptJJJheapq & \fixskelgptfourJJJheapq & \fixskelgptthreeJJJheapq\\

            html & \countfJJJhtml & \fixgptfourbaseJJJhtml & \fixtranscryptJJJhtml & \fixskelgptfourJJJhtml & \fixskelgptthreeJJJhtml\\
            
            mathgen & \countfJJJmathgen & \fixgptfourbaseJJJmathgen & \fixtranscryptJJJmathgen & \fixskelgptfourJJJmathgen & \fixskelgptthreeJJJmathgen\\

            strsim & \countfJJJstrsim & \fixgptfourbaseJJJstrsim & \fixtranscryptJJJstrsim & \cellcolor{lime!30}\fixskelgptfourJJJstrsim & \cellcolor{lime!30}\fixskelgptthreeJJJstrsim\\

            bst-rec & \countfJJJbsdrec & \fixgptfourbaseJJJbsdrec & \cellcolor{lime!30}\fixtranscryptJJJbsdrec & \cellcolor{lime!30}\fixskelgptfourJJJbsdrec & \fixskelgptthreeJJJbsdrec\\

            red-black-tree & \countfJJJrbtree & \fixgptfourbaseJJJrbtree & \fixtranscryptJJJrbtree & \cellcolor{lime!30}\fixskelgptfourJJJrbtree & \fixskelgptthreeJJJrbtree\\
            
            toml & \countfJJJtoml & \fixgptfourbaseJJJtoml & \fixtranscryptJJJtoml & \fixskelgptfourJJJtoml & \fixskelgptthreeJJJtoml\\

            py-evtx & \countfJJJpyevtx & \fixgptfourbaseJJJpyevtx & \fixtranscryptJJJpyevtx & \fixskelgptfourJJJpyevtx & \fixskelgptthreeJJJpyevtx\\
            \midrule
            total & \countfJJJTOTAL & \fixgptfourbaseJJJTOTAL & \fixtranscryptJJJTOTAL & \fixskelgptfourJJJTOTAL & \fixskelgptthreeJJJTOTAL \\
        \bottomrule
    \end{tabular}
\end{table}

%% file: figtabs/tabs/tab4ablation.tex
\begin{table}[h]
\footnotesize
    \centering
    \setlength{\abovecaptionskip}{1ex}
    \caption{Ablation study on two components of \name. The table shows the number of \textbf{\#UserFix} functions in the translated code produced by different versions of \name.}
    \label{tab:ablation}
    \begin{tabular}{l|c|c|c|c}
        \toprule

            \textbf{Programs} & \#Function & $\text{\name}_\text{base}$& $\text{\name}_\text{spec}$& $\text{\name}_\text{spec+chkrfn}$ \\
            
            \midrule
            colorsys & \countfJJJcolorsys & \skelbaseJJJcolorsys & \skelspecJJJcolorsys & \cellcolor{lime!30}\skelchkrfnJJJcolorsys \\
            
            heapq & \countfJJJheapq & \skelbaseJJJheapq & \skelspecJJJheapq & \skelchkrfnJJJheapq \\

            html & \countfJJJhtml & \skelbaseJJJhtml & \skelspecJJJhtml & \skelchkrfnJJJhtml \\
            
            mathgen & \countfJJJmathgen & \skelbaseJJJmathgen & \skelspecJJJmathgen & \skelchkrfnJJJmathgen \\

            strsim & \countfJJJstrsim & \skelbaseJJJstrsim & \cellcolor{lime!30}\skelspecJJJstrsim & \cellcolor{lime!30}\skelchkrfnJJJstrsim \\

            bst-rec & \countfJJJbsdrec & \cellcolor{lime!30}\skelbaseJJJbsdrec & \cellcolor{lime!30}\skelspecJJJbsdrec & \cellcolor{lime!30}\skelchkrfnJJJbsdrec \\

            red-black-tree & \countfJJJrbtree & \skelbaseJJJrbtree & \skelspecJJJrbtree & \cellcolor{lime!30}\skelchkrfnJJJrbtree \\
            
            toml & \countfJJJtoml & \skelbaseJJJtoml & \skelspecJJJtoml & \skelchkrfnJJJtoml \\

            py-evtx & \countfJJJpyevtx & \skelbaseJJJpyevtx & \skelspecJJJpyevtx & \skelchkrfnJJJpyevtx \\
            \midrule
            total & \countfJJJTOTAL & \skelbaseJJJTOTAL & \skelspecJJJTOTAL & \skelchkrfnJJJTOTAL \\
        \bottomrule
    \end{tabular}
\end{table}

%% file: sections/8futurework.tex
\section{Limitations and Open Problems}

We have found \name to be a promising demonstration of the concept of skeletons\tododel{ as a systematic approach} to automate the translation of \py programs up to $2k$ lines of code to \js. In order to scale to even longer programs, we foresee three main challenges that future work may address.
\begin{enumerate}
    \item \textbf{Automated type mapping}. Longer programs often use a broader range of data types. Automated modeling of data types, automated selection of type mapping, as well as the flexibility to change the type mapping for different parts of the program can reduce human effort especially for translating programs involving data types from third-party libraries. 
    \item \textbf{API modeling with opaque side effects}. For APIs that have opaque side effects (such as \code{print(...)}), we currently manually write shims for them, as explained in Section \ref{sec:impl}. Longer programs can have more APIs of this kind or, more severely, APIs that do not have any counterpart in the target language. This makes manual modeling of such APIs tedious and error-prone. Automating such API modeling can be useful to extend \name to translate a larger class of programs.
    \item \textbf{Automatic refactoring for language constructs}. Longer \py programs may use more language features, and some of them (such as multi-inheritance) are unique to the source language, which may result in a high-level program structure that cannot be directly mapped to a valid program in the target language. We think that automatic refactoring of the high-level structure of the source (before the translation starts) might be a reasonable solution, as one can take inspiration from prior experience reports on code migration~\cite{terekhov2000realities}.
\end{enumerate}

Besides the above implementation-level challenges, it is also an open problem to aim for full functional equivalence when translating long programs. \name focuses on test-based equivalence, which does not guarantee equivalence on all possible inputs. A different approach would be to employ formal verification against functional specifications that are, in turn, inferred automatically.

%% file: sections/8relatedwork.tex
\section{Related Work}

This work focuses on automated program translation to produce \tododel{idiomatic} code that satisfies test-based correctness. 
This problem is related to code migration, program synthesis, and specification inference.

\myparagraph{Code Migration} Various approaches have been tried for automated code migration. 
One direction is to build rule-based systems.
The domain-specific language TXL~\cite{CORDY2006190} and the StringTemplate tool~\cite{stringtemplate}, for instance, are general-purpose tools for writing code transformations. 
Developers have built transpilers for specific languages as well, such as Transcrypt---which translates \py to \js~\cite{transcrypt}. 
In theory, such approaches can scale to long programs, but significant development efforts are often needed to build complete enough systems for translating real-world programs.
In the meantime, such rule-based tools often produce non-readable code that emulates the source at the lowest level~\cite{transcrypt,duoglot,c2rust,c2go}.
Another direction besides rule-based systems is to leverage data-driven approaches.
Neural networks can translate code without human-written rules~\cite{chen2018treetotree,lachaux2020unsupervised,6676946}.
With the development of LLMs in recent years, the performance of translating short programs has significant progress~\cite{yin2024rectifier,pan2023stelocoder,chowdhery2023palm,zheng2023survey}. 
Trained on millions of lines of real-world code, they can often produce idiomatic translations with high readability. 
\ignore{LLMs can learn semantic similarities between APIs and operators in different programming languages.}
However, LLMs are error-prone in code translation~\cite{liu2023lost, transmap}.
With the length of the source program increasing, the task quickly exceeds the capacity of LLMs, and the produced code is hardly correct.
Our paper proposes a two-stage solution based on skeleton generation, 
which provides a clean decomposition of the task to allow scalable translation while supporting idiomatic code.
We are aware of concurrent work on the decomposition of translating long programs~\cite{ibrahimzada2024repository,zhang2025skeleton}, but they target partial translations as an aid for human developers and do not aim to pass whole-program tests for the combined translation.

\myparagraph{Program Synthesis} Code translation is also closely related to program synthesis. Program synthesis aims to generate implementation in a target grammar from a specification.
The specification may be in the form of input/output examples~\cite{flashmeta}, logical formulas~\cite{alur2015synthesis}, reference implementations~\cite{verifiedlifting}, inline assertions~\cite{sketch08}, and so on.
Recently, large language models have become another popular avenue for synthesizing programs, and various models have been built or specialized for coding-related tasks~\cite{llmclaude,llmcodellama,llmdeepseek,llmlow,llmstarcoder,codegensynth}.
As for code translation problems, previous work has been applying program synthesis techniques to convert code between different languages.
For example, \citeauthor{verifiedlifting} encodes stencil computations written in Fortran and synthesizes provably correct translations using SMT-solving~\cite{verifiedlifting}.
Such an approach works well for domain-specific languages but is hard to scale to large programs.
\citeauthor{duoglot} synthesizes code translation rules from user snippets and then searches for rule compositions to translate programs~\cite{duoglot}.
However, such an approach has limited scalability due to the exponential search space. 
Our approach explores another attempt to formalize code translation as a synthesis task, where the synthesizer is given the input-output specifications for each placeholder together with the corresponding source fragment as a reference.
The key to its improvements in scalability is a clean decomposition strategy for sub-tasks.
Failures in fragment synthesis can still arise, and future automation techniques may consider leveraging more specialized synthesizers for individual code fragments.

\myparagraph{Specification Inference} Automated inference of specifications has been studied extensively in program analysis and verification~\cite{infbiabduction,infprecond,infpreinfer,infk,infmaxspec,infpermiss,infsec,infbimoving,infdig,infdaikon,infsling}. 
One of the most popular techniques is bi-abduction~\cite{infbiabduction}, which aims to automatically infer pre- and post-conditions of functions for verifying programs in separation logic~\cite{infbimoving,infbishape}. While promising, its success has so far been mostly limited to simple classes of properties rather than functional correctness specifications. A recent technique named Quiver~\cite{infbiquiver} supports inferring functional specifications by guiding abductive inference 
with human-written annotations, thus is able to resolve ambiguity and determine the appropriate level of abstraction for functional specifications. In addition to formal techniques, data-driven approaches to specification inference are also gaining popularity~\cite{infnnc2s,infnnicse,infnnpbt,infnnpldi16,infnnpost,infnnspecgen}. 
While these techniques have the potential to infer full functional specifications, automated validation of the resulting specification remains a challenge~\cite{infnnspecgen}.

%% file: sections/Xconclusion.tex
\section{Conclusion}

We tackle the code translation problem for languages with similar high-level constructs through skeleton generation.
Our approach, \name, translates Python to JavaScript by generating program skeletons with input-output specifications for code fragments derived from tests.
The core idea is to abstract program semantics in both languages into a shared model of communicating processes, capturing observable effects while omitting low-level, language-specific details to enable idiomatic translation.
We also proposed a practical algorithm to fill the skeleton according to the execution order.
We evaluated \name on real-world programs to show its effectiveness.
Several challenges remain, including correctness beyond tests, automating broader data type mappings, modeling of more kinds of APIs, and the support of more language constructs.
Future work addressing these can help improve \name in its ability to translate more complex programs.

\section{Data Availability Statement}

\todoadd{The artifact containing the code and the benchmarks of this paper is available on Zenodo~\cite{skelzen}. The latest version of the artifact can be found here~\cite{githubrepo}.}

%% file: sections/new-A_model.tex
\section{The Execution-Order Translation (EOT) Algorithm}
\label{app:eot}

\input{figtabs/algjit}

As described in Section 4.3, our \name prototype synthesizes the target fragments using the \jitt loop. The precise algorithm for \jitt is shown in Algorithm \ref{alg:jit}.
The input to the algorithm includes (1) the expected execution trace $\rho$ for the translation, which is obtained by applying type mapping on the observable effects of the source program, (2) the skeleton $K^\tgt$ for the target program, and (3) the syntactic content of all the source fragments.
\jitt maintains an append-only set \code{Spec} for each placeholder.
Such \code{Spec} sets serve as counterexamples during synthesis.
At each translation step, \jitt first uses the initial specification to synthesize the content for the fragment if it's missing (lines 7-8).
Then \jitt starts the refinement loop for the fragment.
For each iteration of the specification refinement, \jitt will continue to retry the translation until the fragment satisfies all the specifications in the current counterexample set for the placeholder (lines 12-20). 
When the times of repairing exceed the retry limit, \jitt will ask for help from external, such as asking the human user to provide a fix for the current placeholder (lines 13-14).
To check if the synthesized fragment satisfies the current counterexample set, \jitt runs the incomplete program and compares the obtained $\rho'$ with the expected trace $\rho$ up to the current step (lines 15-18).
Such a comparison can have three potential results.
If the mismatched section occurs in the current counterexample set, it means that the fragment does not satisfy the current counterexample set, and a new fragment needs to be synthesized (repair).
If the fragment satisfies the current counterexample set but still fails on other historical steps for the same fragment (line 21 and line 23), the failed specification (the mismatched section) will be added to the counterexample set (refine), and another iteration of refinement will start.
If there is no mismatched section between $\rho$ and $\rho'$ up to the current step, then it means the fragment satisfies all the historical specifications, and the translation loop can move to the next step (line 22).
The algorithm always terminates since, for each placeholder, there are only a finite number of specifications and a finite number of possible counterexample sets. The algorithm will retry a finite number of times on one counterexample set.
When providing the specifications, we abstract away large objects to reduce the input size for the synthesizer (i.e., an LLM).

\section{More details on the modeling of side effects}
\label{app:side}

As mentioned in Section 4.2, \procint analyzer carefully captures the observable effects of executing code fragments.
Most of the statements, expressions, and APIs appeared in the program, such as \code{x in lst}, \code{sum(lst)}, \code{sort(lst)}, etc.,
can result in state changes in observable objects tracked by \procint analyzer, which will be automatically summarized into the observable effects of the code fragments containing those operations or API calls.
We name these kinds of effects as \emph{transparent} effects.
More than 90\% of the APIs used in the benchmark programs are either pure or result in transparent effects only.

The other category of effects is named \emph{opaque} effects (briefly explained in Section 4.2).
This includes a few of the APIs (<10\%) that have side effects outside of the memory objects tracked by the \procint analyzer.
For instance, \code{random} related APIs have their own internal states preserved across the call (e.g., random seed).
APIs like \code{print} can cause effects on the external environment (write strings to standard output).
The modeling of these APIs for \name requires human effort. The goal here is to model them as process communication messages (to the skeleton process) and such messages should be language-agnostic---they should make sense for both the source and the target language.
To do that, we implement API shims for both \py and \js that intercept the actual API calls to compute the messages that should be sent to the skeleton.
For instance, a call to \code{print} in the source Program will result in a \code{CALL("IO\_WRITE", ..., "<string_to_write>")} effect to the skeleton process.
These messages become part of the semantic requirements and thus need to be explicitly reproduced by the translated program in the exact same order.
For example, the translated program should also send an equivalent \code{CALL("IO\_WRITE", ...)} message to the skeleton at the same relative position in the messages sequence by using JavaScript APIs like \code{console.log}.

\section{The semantics modeling of the skeleton process}
\label{app:sem}

In this section, we formalize the model of communicating processes. The abstract semantic model of the skeleton process $\mathcal{K}$ is illustrated in Fig. \ref{fig:skeletonprocess}.

Compared with the model described in Section 4.2.1, the abstract semantics here include an additional message named \code{start}, which marks the beginning of program execution. The \code{SkelCR} is the syntactic structure of the input program skeleton explained in Section 4.1. At each step, the skeleton process $\mathcal{K}$ receives a message either from a fragment process or from the start point, updates its internal state (including \code{procStack}, \code{procTree}, \code{procObsSet}, and \code{obsObjectStore}), prepares a response message and sends the response to other processes.
For example, when the received message is \code{Call(...)}, the skeleton process initiates a new process using the \code{initNewProcess(...)} function. This function first creates a new fragment process for the given scope, then updates the process stack (\code{procStack}) and process tree (\code{procTree}) to include the new process, and finally returns the new fragment process ID. The skeleton process then reconciles its global object store (\code{obsObjectStore}) with the received object tables \text{CTX} from the fragment process. After updating the state, the skeleton process collects the observable objects ($\text{CTX}_\text{new}$) for the new process using the \code{obtainObservable()} and \code{mayAccessObjects()} functions. The \code{obtainObservable()} function takes the SkelCR syntactic structure and a scope ID as input, and returns the set of observable symbols for the given scope. The \code{mayAccessObjects()} function 
traverses objects reachable from the observable symbols and the previous set of observable objects ($\text{CTX}_\text{old}$), and returns the new set of observable objects $\text{CTX}_\text{new}$. Finally, the skeleton process stores the collected observable objects in \code{procObsSet} and sends $\text{CTX}_\text{new}$ to the new process. Once its task is complete, the skeleton process waits for the next message.
When the message is \code{Ret(...)} received from fragment process, the skeleton process performs a similar update procedure. However, unlike the \code{Call} message, which initializes a new process, the \code{Ret} message signals the termination of the current process and the resumption of the previous one. And rather than generating a new observable object set for the resumed process, the skeleton process updates the existing set using the \code{mayAccessObjects()} function.

Based on this modeling, the semantic for the skeleton $K$ is a sequence of execution steps obtained from the semantic model. Each step consists of one input and one output message. Input messages include \code{Start}, which initiates the execution, as well as \code{Call}, \code{Return}, and \code{Throw} messages from the fragment processes. The output messages can be \code{Init}, \code{Resume}, \code{ResumeThr}, which will be sent to the fragment processes. In contrast, the semantics of each fragment are represented as a set of message sequences, since multiple process instances can correspond to the same fragment. Each message sequence consists of a series of input and output messages. For fragment processes, the input messages can be \code{Init}, \code{Resume}, or \code{ResumeThr}, while the output messages include \code{Call}, \code{Return}, and \code{Throw}. The semantics of the whole program in \name is a sequence derived from the combination of fragment semantics and skeleton semantics. Throughout the whole execution sequence, skeleton semantic and fragment semantic appear in strict alternation. When a process implementation is placed within a program implementation $K, \Gamma$, its contextual semantics are limited and are only a subset of all its possible semantics. Intuitively speaking, the semantics of the program should be valid, i.e., (1) the semantic should start with \code{Start} message and end with stop execution, (2) each fragment process should start will \code{Init} and end with \code{Ret} or \code{Throw}, and (3) each step should be compatible with the step before and after it. \ignore{All the satisfying semantic is a set ${\sf Sem}_{K, \Gamma}(g) \subset {\sf Sem}(g).$}

After obtaining the semantics of the source program, \name prototype uses type mapping $\mathcal{F}$ to convert the semantics of the source code fragments to the semantic requirements for the translated fragments. The type mapping applied by \name prototype is shown in Fig. \ref{fig:appendix_mapping}.

\begin{figure}[!htbp]
    \centering
    \footnotesize
    \begin{itemize}
        \item Input: an input message $\textbf{Input}$.
        \item Output: an response message $\textbf{Output}$.
        \item match $\text{Input}$ with:
        \begin{itemize}

            \item \textbf{Input} is START(SkelCR) \graytext{// Start Execution.}
            \begin{itemize}
                \item $\text{procStack}, \text{procTree}, \text{procObsSet}, \text{obsObjectStore} \leftarrow \emptyset, \emptyset, \emptyset, \emptyset$ \graytext{// Initialize the internal states.}
                \item $\text{Id}^\text{Proc}_\text{new}, \text{procStack}, \text{procTree}  \leftarrow  \text{initNewProcess}(\text{SkelCR}, \text{Id}^\text{Scope}_\text{global}, null, null, \text{procStack}, \text{procTree})$
                \item $\text{Symbol}_\text{obs} = \text{obtainObservable($\text{SkelCR}, \text{Id}^\text{Scope}_\text{global}$)}$
                \item $\text{CTX}_\text{new} = \text{mayAccessObjects($\text{Symbol}_\text{obs}$, $\text{obsObjectStore}$, null)}$
                \item $\text{procObsSet}[\text{Id}^\text{Proc}_\text{new}] = \text{CTX}_\text{new}$
                \item Send \textbf{Output} Init($\text{CTX}_\text{new}$) to $\text{Id}^\text{Proc}_\text{new}$ process and wait for next message.
            \end{itemize}

            \item \textbf{Input} is $\text{Call}(\text{CTX}, \text{Closure}(\text{Id}^\text{Scope}, \text{Id}^\text{Proc}), \text{ARGS})$  \graytext{// Initialize a new process.}
            \begin{itemize}
                \item $\text{Id}^\text{Proc}_\text{new}, \text{procStack}, \text{procTree}  \leftarrow  \text{initNewProcess}(\text{SkelCR}, \text{Id}^\text{Scope}, \text{Id}^\text{Proc}, \text{ARGS}, \text{procStack}, \text{procTree})$
                \item $\text{Symbol}_\text{obs} = \text{obtainObservable($\text{SkelCR}, \text{Id}^\text{Scope}$)}$
                \item $\text{obsObjectStore} \leftarrow \text{reconcileState}(\text{obsObjectStore}, \text{CTX})$
                \item $\text{CTX}_\text{new} = \text{mayAccessObjects($\text{Symbol}_\text{obs}$, $\text{obsObjectStore}$, null)}$
                \item $\text{procObsSet}[\text{Id}^\text{Proc}_\text{new}] = \text{CTX}_\text{new}$
                \item Send \textbf{Output} Init($\text{CTX}_\text{new}$) to $\text{Id}^\text{Proc}_\text{new}$ process and wait for next message.
            \end{itemize}

            \item \textbf{Input} is $\text{Ret}(\text{CTX}, \text{Id}^\text{obj})$ \graytext{// Stop the current process and resume the previous process.}
            \begin{itemize}
                \item $\text{Id}^\text{Proc}_\text{previous}, \text{Id}^\text{Scope}_\text{previous}, \text{procStack} \leftarrow \text{stopCurrentProcess}(\text{procStack})$
                \item \textbf{If} $\text{Id}^\text{Proc}_\text{Previous} = null$ \textbf{then}: \textbf{Stop the execution}.
                \item $\text{Symbol}_\text{obs} = \text{obtainObservable($\text{SkelCR}, \text{Id}^\text{Scope}$)}$
                \item $\text{obsObjectStore} \leftarrow \text{reconcileState}(\text{obsObjectStore}, \text{CTX})$
                \item $\text{CTX}_\text{old} = \text{procObsSet}[\text{Id}^\text{Proc}_\text{previous}]$
                \item $\text{CTX}_\text{new} = \text{mayAccessObjects($\text{Symbol}_\text{obs}$, $\text{obsObjectStore}$, $\text{CTX}_\text{old}$)}$
                \item $\text{procObsSet}[\text{Id}^\text{Proc}_\text{previous}] = \text{CTX}_\text{new}$
                \item Send \textbf{Output} Resume($\text{CTX}_\text{new}$, $\text{Id}^\text{obj}$) to $\text{Id}^\text{Proc}_\text{Previous}$ process and wait for next message.
            \end{itemize}

            \item  \textbf{Input} is $\text{Throw}(\text{CTX}, \text{Id}^\text{obj})$ \graytext{// Stop the current process and resume the previous process.}
            \begin{itemize}
                \item $\text{Id}^\text{Proc}_\text{previous}, \text{Id}^\text{Scope}_\text{previous}, \text{procStack} \leftarrow \text{stopCurrentProcess}(\text{procStack})$
                \item \textbf{If} $\text{Id}^\text{Proc}_\text{Previous} = null$ \textbf{then}: \textbf{Stop the execution}.
                \item $\text{Symbol}_\text{obs} = \text{obtainObservable($\text{SkelCR}, \text{Id}^\text{Scope}$)}$
                \item $\text{obsObjectStore} \leftarrow \text{reconcileState}(\text{obsObjectStore}, \text{CTX})$
                \item $\text{CTX}_\text{old} = \text{procObsSet}[\text{Id}^\text{Proc}_\text{previous}]$
                \item $\text{CTX}_\text{new} = \text{mayAccessObjects($\text{Symbol}_\text{obs}$, $\text{obsObjectStore}$, $\text{CTX}_\text{old}$)}$
                \item $\text{procObsSet}[\text{Id}^\text{Proc}_\text{previous}] = \text{CTX}_\text{new}$
                \item Send \textbf{Output} ResumeThr($\text{CTX}_\text{new}$, $\text{Id}^\text{obj}$) to $\text{Id}^\text{Proc}_\text{Previous}$ process and wait for next message.
            \end{itemize}

        \end{itemize}

    \end{itemize}

    \footnotesize
    \caption{The semantic model of the skeleton process $\mathcal{K}$.}
    \label{fig:skeletonprocess}
\end{figure}

\begin{figure}
    \centering
    \footnotesize
    \[
    \begin{array}{lcl}
        \mathcal{F}: \text{Python Types} & \to & \text{JavaScript Types} \\
        \mathcal{F}(\texttt{Int}) & = & \texttt{Number} \\
        \mathcal{F}(\texttt{Float}) & = & \texttt{Number} \\
        \mathcal{F}(\texttt{Str}) & = & \texttt{String} \\
        \mathcal{F}(\texttt{Bool}) & = & \texttt{Boolean} \\
        \mathcal{F}(\texttt{NoneType}) & = & \texttt{Null} \\
        \mathcal{F}(\texttt{Bytes}) & = & \texttt{Uint8Array} \\
        \mathcal{F}(\texttt{List}[\tau_1, \tau_2, \dots, \tau_n]) & = & \texttt{Array}[\mathcal{F}(\tau_1), \mathcal{F}(\tau_2), \dots, \mathcal{F}(\tau_n)] \\
        \mathcal{F}(\texttt{Tuple}[\tau_1, \tau_2, \dots, \tau_n]) & = & \texttt{Array}[\mathcal{F}(\tau_1), \mathcal{F}(\tau_2), \dots, \mathcal{F}(\tau_n)] \\
        \mathcal{F}(\texttt{Set}\{\tau_1, \tau_2, \dots, \tau_n\}) & = & \texttt{Set}\{\mathcal{F}(\tau_1), \mathcal{F}(\tau_2), \dots, \mathcal{F}(\tau_n)\} \\
        \mathcal{F}(\texttt{Dict}\{\tau_1^{key}: \tau_1^{val}, \dots, \tau_n^{key}: \tau_1^{val}\}) & = & \texttt{Object}\{\mathcal{F}(\tau_1^{key}): \mathcal{F}(\tau_1^{val}), \dots, \mathcal{F}(\tau_n^{key}): \mathcal{F}(\tau_n^{val})\}\\
        \mathcal{F}(\texttt{Closure}) & = & \texttt{Closure} \\
    \end{array}
    \]
    \caption{The type mapping used in \name prototype (detailed mapping of data values is omitted).}
    \label{fig:appendix_mapping}
\end{figure}

\section{\py to \decocr: Source Normalization}
\label{app:norm}

In the main paper, we explained how basic Python code corresponds to \decocr. The example demonstrates basic features, including closure declarations, local and non-local variable declarations, sequential arguments passing, etc. Although these features have a direct correspondence to \decocr, themselves alone will be insufficient if we want to translate realistic Python programs. To support more language features, we implemented a source normalizer to rewrite a set of richer Python language features into a smaller language subset that has a straightforward mapping to \decocr. Table \ref{tab:reduction} shows at a high level a list of language feature rewriting strategies we implemented.
A caveat of such normalization is that it breaks the native support of operator overloading in Python, such as the class method \code{def __eq__(self)} that will interact with the \code{==} operator. However, it is not a major problem in our benchmarks. For a handful of places that involve calling overloaded operators, we perform an additional dynamic analysis specifically on operators to locate them and replace them into \code{__eq__} calls.

\begin{myverbbox}[\tiny]{\codeEgA}
class MyClass:
  def __init__(self):
    self.myvar = 3
  def update(self):
    self.myvar += 3
\end{myverbbox}

\begin{myverbbox}[\tiny]{\codeEgAnorm}
def MyClass():
  def __init__():
    class_var.myvar = 3
  def update():
    class_var.myvar += 3
  class_var = SkelClass('MyClass')
  class_var.update = update
  __init__()
  return class_var
\end{myverbbox}

\begin{myverbbox}[\tiny]{\codeEgC}
class Car: ...
  def __init__(self, brand):
     ...
class ECar(Car): 
  def __init__(self, brand, battery):
    super().__init__(brand)
    self.battery = battery
    
\end{myverbbox}

\begin{myverbbox}[\tiny]{\codeEgCnorm}
def Car(brand):
  ...
def ECar(brand, battery):
  def __init__(brand, battery):
    class_var.battery = battery
  class_var = Car(brand)
  __init__(brand, battery)
  return class_var
\end{myverbbox}

\begin{myverbbox}[\tiny]{\codeEgD}
def greet(name, age):
  print(f"Hello, {name}!...")
greet(age=30, name="bob")
greet(name="alice", age=25)
\end{myverbbox}

\begin{myverbbox}[\tiny]{\codeEgDnorm}
def greet(name, age): 
  print(...)
greet("bob", 30) # modified
greet("alice", 25) # modified
\end{myverbbox}

\begin{myverbbox}[\tiny]{\codeEgE}
def greet(name, age=25): 
  print(...)
greet("bob", 30)
greet("alice")
\end{myverbbox}

\begin{myverbbox}[\tiny]{\codeEgEnorm}
def greet(name, age): # modified
  print(...)
greet("bob", 30)
greet("alice", 25) # modified
\end{myverbbox}

\begin{myverbbox}[\tiny]{\codeEgF}
def deco_uppercase(func):
  def wrapper():
    return func().upper()
  return wrapper
@deco_uppercase
def greet():
  return "hello"
\end{myverbbox}

\begin{myverbbox}[\tiny]{\codeEgFnorm}
def deco_uppercase(func):
  def wrapper():
    return func().upper()
  return wrapper
def _greet():
  return "hello"
greet = deco_uppercase(_greet)
\end{myverbbox}

\newcommand{\mycentered}[1]{\begin{tabular}{l}#1 \end{tabular}}
\begin{table}[h]
    \scriptsize
    \centering
    \caption{Rewriting Strategies for Python Language Features to a Subset Compatible with \decocr}
    \label{tab:reduction}
    \begin{tabular}{p{0.14\linewidth} p{0.27\linewidth} p{0.22\linewidth} p{0.27\linewidth}}
        \toprule
        \textbf{Language Feature} & \textbf{~Code Example} & \textbf{Rewriting Strategy} & \textbf{~Code Example} \\
        \midrule
        \mycentered{Class Declarations} & \mycentered{\codeEgA} & \mycentered{Rewriting classes methods to \\ closures, and class constructors \\ return a dict-like object.} & \mycentered{\codeEgAnorm} \\
        \midrule
        \mycentered{Class Inheritance} & \mycentered{\codeEgC} & \mycentered{Rewriting inheritance into\\calls to normalized class\\constructor of base class} & \mycentered{\codeEgCnorm} \\
        \midrule
        \mycentered{Keyword Arguments} & \mycentered{\codeEgD} & \mycentered{Re-ordering arguments at\\callsites to sequential \\argument passing} & \mycentered{\codeEgDnorm} \\
        \midrule
        \mycentered{Default Arguments} & \mycentered{\codeEgE} & \mycentered{Inserting default arguments\\into callsites} & \mycentered{\codeEgEnorm} \\
        \midrule
        \mycentered{Decorators} & \mycentered{\codeEgF} & \mycentered{Rewriting the decorator into a\\call that returns a closure} & \mycentered{\codeEgFnorm} \\
        \bottomrule
    \end{tabular}    
\end{table}

\section{Prompt Structure used in evaluation}
\label{app:prompt}

\begin{figure}[t]
    \centering
    \includegraphics[width=0.85\linewidth]{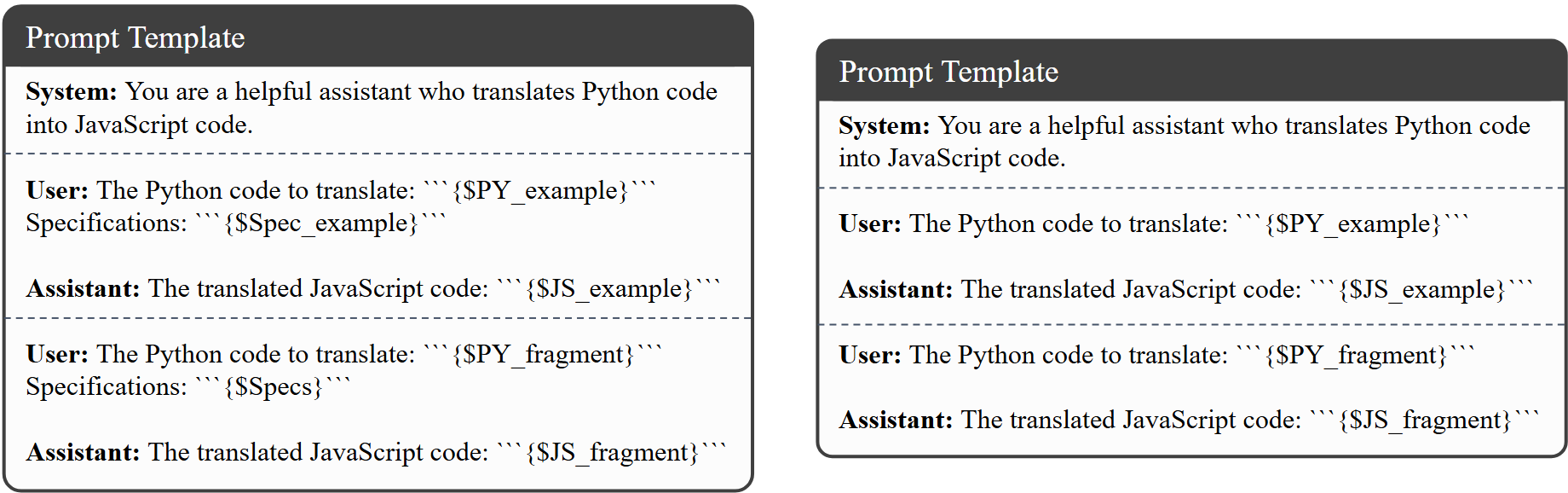}
    \caption{The prompt structure for \name prototype (left) and the prompt structure for baseline (right) in evaluation. Each of them is composed of (1) a system prompt, (2) one-shot example, and (3) real input.}
    \label{fig:appendix_prompt}
\end{figure}

The prompt structure used for \name prototype during evaluation is shown on the left of Figure \ref{fig:appendix_prompt}.
The prompt structure for the baseline approach is the same, except there are no specifications.

\section{Failed cases of Synthesizers}
\label{app:fail}
During translating our benchmarks of 9 programs, \name prototype correctly translates 95\% of the functions in total. For the remaining 5\% cases where LLMs cannot produce correct translations satisfying the specifications. We further analyze their root cause. A large group of errors is caused by \emph{the mismatch of behaviors of similar APIs and operators} across languages. 
In these cases, even if provided with counterexamples, LLMs tend to trust those APIs and operators, and new translations keep using them wrongly. Most of the errors can be attributed to this group. It's hard to determine the reason for the remaining errors, including wrongly changing the structure of the code, missing part of the code, using APIs and libraries that don't exist, etc.

%% file: figtabs/algjit.tex
\begin{algorithm}[t]
\caption{Execution-Order Translation Algorithm}
\label{alg:jit}
\begin{algorithmic}[1]
    \item  \textbf{Input:} $\rho$ \graytext{// the expected trace for the translation "$\mathcal{K}  \myarrow{\text{Input}} \textcolor{violet}{P_{i}} \myarrow{\text{Output}} \mathcal{K} ...$"}
    \State $\skl^\tgt, \{g^{\text{src}}_0, g^{\text{src}}_1, ... ,g^{\text{src}}_n\}$, \text{limit}
    \item \textbf{Output:} $\{g^{\text{tgt}}_0, g^{\text{tgt}}_1, ... ,g^{\text{tgt}}_n\}$

    \State $\text{Spec}_0, \text{Spec}_1, ..., \text{Spec}_n \leftarrow \{\}$ \graytext{// One counterexample set for one placeholder}
    \State $g^{\text{tgt}}_0, g^{\text{tgt}}_1, ... ,g^{\text{tgt}}_n \leftarrow \text{Null}$
    
    \For {$(\mathcal{K}, \text{Input}, \textcolor{violet}{P}, \text{Output}) \gets \rho$} \graytext{// Each step is a small section "$\mathcal{K}  \myarrow{\text{Input}} \textcolor{violet}{P} \myarrow{\text{Output}}$" of $\rho$}

    \State $\text{Id} \gets \text{getFragment}(\textcolor{violet}{P})$ \graytext{// Get the Id of the corresponding fragment}

    \If {$g^{\text{tgt}}_\text{Id} = \text{Null}$}\label{line:misblock1} \graytext{// Check if $g^{\text{tgt}}_\text{Id}$ is a missing fragment}
        \State $g^{\text{tgt}}_\text{Id} \leftarrow \text{fragSynth}(\{\text{Input}, \text{Output}\}, g^{\text{src}}_\text{Id})$ \graytext{// Synthesize only with initial specification} \label{line:misblock2}
    \EndIf
    
    \While {$\text{True}$} \graytext{// Refinement loop}
    \State {$\text{Mismatch} \leftarrow \text{Null}$}
    \State {$\text{count} \leftarrow \text{0}$}

    \While {\text{True}} \graytext{// Repair loop}
        \If {$\text{count} > \text{limit}$}
        \State $g_\text{Id}^\text{tgt} \leftarrow \text{askExternalAid}(\text{Spec}_\text{Id}, g^{\text{src}}_\text{Id})$
        \EndIf
        
        \State $\rho' \leftarrow {\procint}(K^\tgt, \{g^{\text{tgt}}_0, g^{\text{tgt}}_1, ... ,g^{\text{tgt}}_n\})$\label{line:get trace py}

        \State $\text{Mismatch} \leftarrow \text{getMismatch}(\rho, \rho')$
        \If {$\text{Mismatch} \notin \text{Spec}_\text{Id}$} \label{line:step check} \graytext{// Check if $g_\text{Id}^\text{tgt}$ satisfy the current counterexample set}
            \State $\textbf{break}$ \graytext{// Satisfy the current counterexample set}
        \EndIf
    
        \State $g^{\text{tgt}}_\text{Id} \leftarrow \text{fragSynth}(\text{Spec}_\text{Id}, g^{\text{src}}_\text{Id})$  \graytext{// Synthesize with counterexamples}
        
        \State $\text{count} \leftarrow \text{count} + 1$

    \EndWhile
    \If {$\text{Mismatch} = \text{Null}$} \label{line:step check} \graytext{// Check whether the fragment fails on other specifications}
            \State $\textbf{break}$ \graytext{// Pass the current step}
    \EndIf
        
    \State $\text{Spec}_\text{Id} \leftarrow \text{Spec}_\text{Id} \bigcup \{\text{Mismatch}\}$ \graytext{// Refine the counterexample set} 
    \EndWhile

    \EndFor
    
    \State $\textbf{Return} ~ \{g^{\text{tgt}}_0, g^{\text{tgt}}_1, ... ,g^{\text{tgt}}_n\}$
\end{algorithmic}
\end{algorithm}